\documentclass[final,5p,times,twocolumn]{elsarticle}
\pdfoutput=1
\usepackage{graphicx}
\usepackage{subfig}
\usepackage{amssymb}
\usepackage{lineno}
\usepackage{pslatex}
\usepackage{amsmath}
\usepackage[colorinlistoftodos]{todonotes}
\usepackage{hyperref}
\journal{Journal}
\makeatletter
\def\ps@pprintTitle{%
 \let\@oddhead\@empty
 \let\@evenhead\@empty
 \def\@oddfoot{}%
 \let\@evenfoot\@oddfoot}
\makeatother
\usepackage{pdfpages}
\begin{document}

\begin{frontmatter}

\title{Modelling compensated antiferromagnetic interfaces with MuMax$^3$}
\author[gent]{Jonas De Clercq \corref{cor}}
\ead{Jonas.DeClercq@ugent.be}
\author[gent]{Jonathan Leliaert}
\author[gent]{Bartel Van Waeyenberge}

\address[gent]{DyNaMat, Department of Solid State Sciences, Ghent University, Krijgslaan 281 S1, B-9000 Gent, Belgium}

\cortext[cor]{Corresponding author}

\begin{abstract}  \\
We show how compensated antiferromagnetic spins can be implemented in the micromagnetic simulation program MuMax$^3$. We demonstrate that we can model spin flop coupling as a uniaxial anisotropy for small canting angles and how we can take into account the exact energy terms for strong coupling between a ferromagnet and a compensated antiferromagnet. We also investigate the training effect in biaxial antiferromagnets and reproduce the training effect in a polycrystalline IrMn/CoFe bilayer.
\end{abstract}
\end{frontmatter}

\section{Introduction}
\label{S:1} 
Although the first clues about the interaction between an antiferromagnet (AFM) and a ferromagnet (FM) were already found 60 years ago with the discovery of exchange bias by Meiklejohn and Bean\cite{Meiklejohn_Bean}, still not all details are fully understood today. The shift of the hysteresis loop when a FM/AFM bilayer is cooled in an external field below the N\'eel temperature $T_\mathrm{N}$ and the increased coercivity of the loop were explained by considering uncompensated AFM spins at the interface, i.e. only one of the AFM sublattices couples to the FM. Theoretical calculations have shown that the surface energy density $J_\mathrm{I}$ of this interface interaction is one or two orders of magnitude larger than the experimentally measured value.\cite{j_eb_ideal} This led to the conclusion that only a small fraction of the interfacial AFM spins contribute to the exchange bias in most systems. The other spins do not couple or rotate together with the FM due to a locally reduced anisotropy or a stronger interfacial coupling.\cite{radu}

Besides an increase in coercivity and a shift of the hysteresis loop, in most polycrystalline bilayers also a training effect can be observed, i.e. the bias field $B_\mathrm{eb}$ and the coercivity $B_\mathrm{c}$ decrease for an increasing number of hysteresis cycles $n$. This training effect can have two contributions: thermal and athermal training. Thermal training, which is generally much smaller than athermal training, happens for $n \geq 1$ and is rather well understood. Thermal fluctuations lead to a depinning of the frozen uncompensated spins at the interface and thus to a decrease in the bias field. Theory\cite{thermal} shows that thermal training follows a power law given by $B_\mathrm{eb} (n) \propto \frac{1}{\sqrt{n}}$, which has been confirmed extensively by experimental data.\cite{thermal_0,thermal_1,thermal_2} Athermal training predominantly happens in the first hysteresis loop and often leads to an asymmetry in the first reversal of the FM towards negative saturation. It results from the frustrated metastable state of rotatable AFM spins after field cooling. 

In a previous paper\cite{eigen_exchange_bias} we have shown that we were able to model exchange bias and training effects due to frozen and rotatable uncompensated AFM spins using MuMax$^3$\cite{MuMax3}, which is an open source micromagnetic simulation package that was primarily designed to study static and dynamic effects in ferromagnets. This GPU-accelerated software also allows for an easy implementation of the microstructure e.g. to divide a micrometer sized sample into small grains by using a Voronoi tesselation.\cite{voronoi} 

When most of the AFM spins at the interface are compensated, i.e. if both sublattices of the AFM couple equally to the FM layer, the magnetic system will try to minimize its total energy by canting the AFM spins towards the FM.\cite{spin_flop1,spin_flop2,SAF} This second order magnetic interaction\cite{second_order}, called spin flop coupling, produces a small net magnetic moment in the AFM and an increased coercivity\cite{schulthess_spin_flop}. Micromagnetic simulations\cite{schulthess_spin_flop} and theoretical considerations have shown however that for an AFM with uniaxial anisotropy this does not lead to exchange bias. Although the origin of this spin flop coupling is well understood, there are still a lot of unanswered questions, e.g. how is this increase in coercivity related to fundamental parameters and how and when can training effects result from compensated AFM spins?

In this paper we will demonstrate how we can model spin flop coupling and training effects of a compensated AFM interface using MuMax$^3$. In the next section we will consider a simple model of an AFM with a strong uniaxial anisotropy and show that for small canting angles, spin flop coupling can be included by adding a custom energy density term $\epsilon_\mathrm{sf}$ and a custom field term $\vec{B}_\mathrm{sf}$ in MuMax$^3$. Although this implementation is very efficient as only the ferromagnet needs to be taken into account, it has several drawbacks, e.g. these approximations are no longer valid for strong coupling parameters $J_\mathrm{I}$ and training effects cannot be produced using this approach. 

In the third section we will show how MuMax$^3$ can also simulate large canting angles by modelling a compensated AFM interface, existing of 2 atomic AFM sublattices, by adding 2 extra layers in the micromagnetic model. Finally, the influence of spin flop coupling on the Landau state in square magnetic nanostructures will be compared to experimental data and the case of training effects in an AFM with biaxial anisotropy will be discussed. 

Having already shown that we can model uncompensated spins in a previous paper\cite{eigen_exchange_bias}, we are now able to demonstrate for the first time how a mixed interface, containing compensated as well as uncompensated AFM spins can be taken into account in one micromagnetic model. In this way we can offer a good description of the static effects due to the interface coupling between a FM and an AFM. Also the coupling between synthetic antiferromagnets\cite{SAF2} and a FM layer can be modelled using this approach. 

\section{Spin flop coupling for small canting angles}
\label{S:2}
\begin{figure}[htb]
\centering
\includegraphics[width=0.3\textwidth]{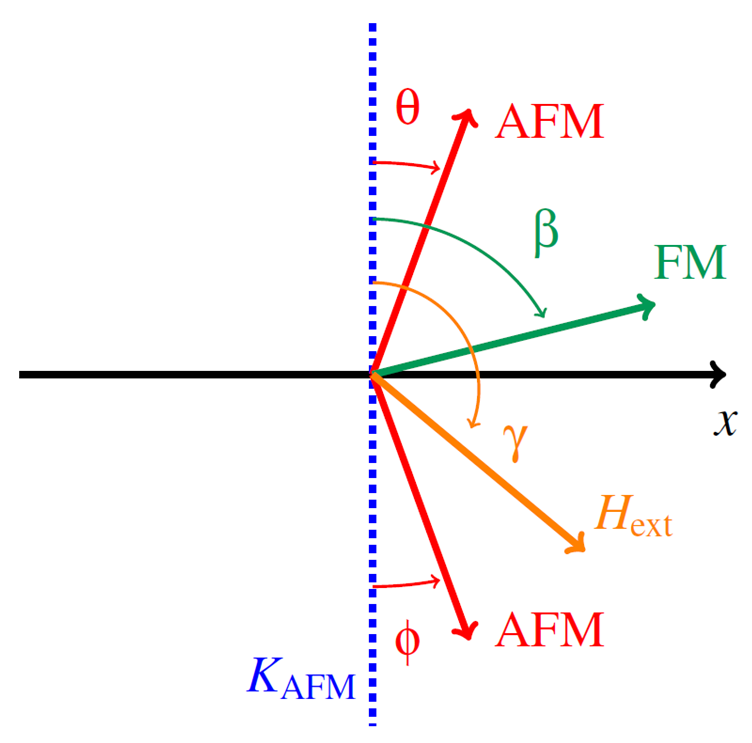}
\caption{Definition of the AFM canting angles $\theta$ and $\phi$, the FM angle $\beta$ and the angle $\gamma$ of the external field $H_\mathrm{ext}$ with respect to the AFM uniaxial anisotropy axis $K_\mathrm{AFM}$.}
\label{fig:spin_flop_theory} 
\end{figure}

As a toy model to describe spin flop coupling with a uniaxial AFM, we follow the macrospin approach where each magnetic subsystem (e.g. the FM and AFM sublattices) is assumed to be in a uniform state and can be described by a single magnetization vector. 
\\
We consider an infinite in-plane magnetized FM film (with thickness $t_\mathrm{FM}$ and saturation magnetization $M_\mathrm{FM}$) coupled to a compensated antiferromagnet. The total surface energy density $\sigma$ of this system can be written as
\begin{align}
\sigma = &- \mu_0 M_\mathrm{FM}t_\mathrm{FM}H_\mathrm{ext} \cos(\gamma - \beta) + \epsilon_\mathrm{K}(\beta) t_\mathrm{FM} \nonumber \\
&+ \sigma_\mathrm{AFM}(\beta, \phi, \theta)
\label{eq:exact_energy}
\end{align}
where the function $\epsilon_\mathrm{K}(\beta)$ represents the anisotropy energy density of the FM and $\beta$ and $\gamma$ denote the angles that the FM and the external field $H_\mathrm{ext}$ make with the AFM anisotropy axis, respectively. We assume that the FM has a uniform uniaxial anisotropy perpendicular to the N\'eel axis, so $\epsilon_\mathrm{K} (\beta) = K_\mathrm{FM} \cos^2(\beta)$. The function $\sigma_\mathrm{AFM}(\beta, \phi, \theta)$, which describes the interaction of the AFM (total thickness $t_\mathrm{AFM}$) with the FM and the internal interaction between the AFM macrospins, is given by
\begin{align}
\label{eq:s_afm}
\sigma_\mathrm{AFM}(\beta, \phi, \theta)= &- J_\mathrm{I} \cos(\beta - \theta) + J_\mathrm{I}\cos(\beta + \phi) \nonumber \\
&- K_\mathrm{AFM} t_\mathrm{AFM} \left[ \cos^2(\theta)  + \cos^2(\phi)\right]\nonumber\\ &- \delta t_\mathrm{AFM} \cos(\theta + \phi)
\end{align}
where $J_\mathrm{I}$ is the coupling constant (surface energy density) between the FM and an AFM macrospin, $K_\mathrm{AFM}$ is the anisotropy constant of a sublattice and $\delta$ the energy density linked to the mutual interaction between the 2 AFM sublattices. The definition of the angles $\phi$ and $\theta$, which the 2 AFM macrospins make with their anisotropy axes, is shown in figure \ref{fig:spin_flop_theory}.
\\
As only the term $\sigma_\mathrm{AFM}$ in $\sigma$ depends on the angles $\phi$ and $\theta$, it suffices to calculate the derivatives of $\sigma_\mathrm{AFM}$ towards $\phi$ and $\theta$ in order to minimize the total energy of the AFM. This leads to equations for $\theta$ and $\phi$ as a function of the angle $\beta$.
Assuming small canting angles, we can expand the function $\sigma_\mathrm{AFM}$ up to second order in $\theta$ and $\phi$ and so we can approximate the AFM energy density as 
\begin{align} \label{eq:linearised}
\sigma_\mathrm{AFM}  \approx &- J_\mathrm{I} \left[ \left( 1- \frac{\theta^2}{2}\right) \cos(\beta) + \theta \sin(\beta) \right] \nonumber \\
&+ J_\mathrm{I} \left[ \left( 1- \frac{\phi^2}{2}\right) \cos(\beta) - \phi \sin(\beta) \right] \nonumber \\
&+ K_\mathrm{AFM} t_\mathrm{AFM} \left( \theta^2 + \phi^2 \right) + \frac{\delta t_\mathrm{AFM}}{2}\left( \theta + \phi \right)^2
\end{align}
leaving out constant energy terms. Minimizing this energy density by calculating $\frac{\partial \sigma_\mathrm{AFM}}{\partial \theta}$ and $\frac{\partial \sigma_\mathrm{AFM}}{\partial \phi}$, we find an expression for $\theta$ and $\phi$ as a function of the FM angle $\beta$
\begin{align} \label{eq:theta_phi}
&\theta(\beta) = \frac{J_\mathrm{I} \sin(\beta) \left[ - J_\mathrm{I}\cos(\beta) +  2 K_\mathrm{AFM} t_\mathrm{AFM} \right]}{4 t_\mathrm{AFM}^2 K_\mathrm{AFM} (K_\mathrm{AFM} + \delta) - J^2_\mathrm{I}\cos^2(\beta)}
\\
&\phi(\beta) = \frac{J_\mathrm{I} \sin(\beta) \left[  J_\mathrm{I} \cos(\beta) +  2 K_\mathrm{AFM} t_\mathrm{AFM} \right]}{4 t_\mathrm{AFM}^2 K_\mathrm{AFM} (K_\mathrm{AFM} + \delta) - J^2_\mathrm{I}\cos^2(\beta)}
\end{align}
These 2 equations show that for $\beta = \frac{\pi}{2}$ the angles of the AFM macrospins are symmetrical around the FM direction $\beta$, i.e. $\theta = \phi$. After substituting these angles $\theta(\beta)$ and $\phi(\beta)$, which minimize the AFM energy density $\sigma_\mathrm{AFM}$, in equation \ref{eq:linearised}, we find that 
\begin{equation}
\sigma_\mathrm{AFM} (\beta) \approx - \frac{ \kappa \sin^2(\beta)}{1 - \frac{\kappa \cos^2(\beta)}{2 t_\mathrm{AFM} K_\mathrm{AFM}}}
\end{equation}
where we have defined the constant $\kappa = \frac{J^2_\mathrm{I}}{2 t_\mathrm{AFM} (K_\mathrm{AFM} + \delta)}$. Retaining only the lowest order approximation in $\frac{\kappa}{2 t_\mathrm{AFM} K_\mathrm{AFM}}$  in the case of low coupling, one finally obtains that
\begin{equation} \label{eq_sf}
\sigma_\mathrm{AFM} (\beta)  \approx \kappa \cos^2(\beta)
\end{equation}
This equation describes the spin flop coupling for small canting angles, i.e. the energy of the antiferromagnet is minimal when the ferromagnet is perpendicular $ \left( \beta = \frac{\pi}{2} \right)$ to the AFM N\'eel vector  and so in this case $\theta = \phi = \frac{J_\mathrm{I}}{2 t_\mathrm{AFM} (K_\mathrm{AFM} + \delta)}$. 
\\
The total surface energy density can then be written as 
\begin{align}
\sigma = &- \mu_0 M_\mathrm{FM}t_\mathrm{FM}H_\mathrm{ext} \cos(\gamma - \beta) + \left( K_\mathrm{FM} t_\mathrm{FM}  + \kappa\right) \cos^2(\beta)
\end{align}
and the switching field $B_\mathrm{c}$ for $\gamma = \frac{\pi}{2}$ is given by 
\begin{align} \label{eq:switching}
B_\mathrm{c} = \frac{2}{M_\mathrm{FM}} \left[ K_\mathrm{FM} + \frac{J^2_\mathrm{I}}{2 t_\mathrm{AFM} t_\mathrm{FM} \left( K_\mathrm{AFM} + \delta \right)} \right] 
\end{align}
This shows that spin flop coupling only leads to a renormalization of the uniaxial anisotropy constant $K_\mathrm{FM}$ and thus to an enhanced coercivity for a hysteresis loop measured perpendicular to the N\'eel vector of the AFM. 
\\
Given the parameters $J_\mathrm{I}$, $K_\mathrm{AFM}$ and $\delta$, one can implement spin flop coupling in MuMax$^3$ by adding a custom energy density $\epsilon_\mathrm{sf}$ and a corresponding effective field $\vec{B}_\mathrm{sf}$, defined as
\begin{align} \label{eq:Ku}
\epsilon_\mathrm{sf} &:= - \frac{1}{2} \vec{M}_\mathrm{FM} . \vec{B}_\mathrm{sf} = \frac{\kappa}{t_\mathrm{FM}} \cos^2(\beta) \\
\vec{B}_\mathrm{sf} &= - \frac{2\kappa}{M_\mathrm{FM} t_\mathrm{FM}} (\vec{u} .\vec{m})\vec{u}
\end{align}
with $\vec{m}$ the normalized magnetization vector of the FM and $\vec{u}$ the N\'eel vector, i.e. the anisotropy direction of the AFM. As expected, this spin flop model does not produce exchange bias. 
\\
In case the 2 AFM macrospins couple with different strengths to the FM, one can approximate the exact energy terms again for small canting angles and show that this leads to exchange bias as a first order effect, given by $\sigma_\mathrm{AFM} (\beta) = -\left[J_{\mathrm{I},\theta} - J_{\mathrm{I},\phi}\right]\cos(\beta)$.

\section{Full micromagnetic description of spin flop coupling}
\label{S:3}
\subsection{Micromagnetic model}\label{mm_model}

In a micromagnetic approach, the atomic magnetic moments and their quantum mechanical interactions are averaged over a nanometer length scale, which is sufficiently small to resolve magnetic structures like domain walls. This approach has been used for many years to model ferromagnets, where the exchange interaction does not allow for sharp changes in the magnetization. In an AFM where the magnetic moments alternate direction on consecutive atomic sites, this approach would result in a zero net magnetization. However, at the micromagnetic scale, the two atomic sublattices of an AFM can be considered as two separate, smoothly varying ferromagnetically ordered lattices which are antiferromagnetically coupled and coincide in space. 

Disregarding the atomic scale of these sublattices means that the dynamics relevant on this length scale (e.g. AFM spin waves) are lost, just like the high frequency part of the spin wave spectrum is also lost in a FM model when the local variations are averaged out. However, the static interaction between the 2 sublattices is not lost as the exchange interaction between the sublattices is taken into account in the micromagnetic model.

\begin{figure}[htb]
\centering
\includegraphics[width=0.48\textwidth]{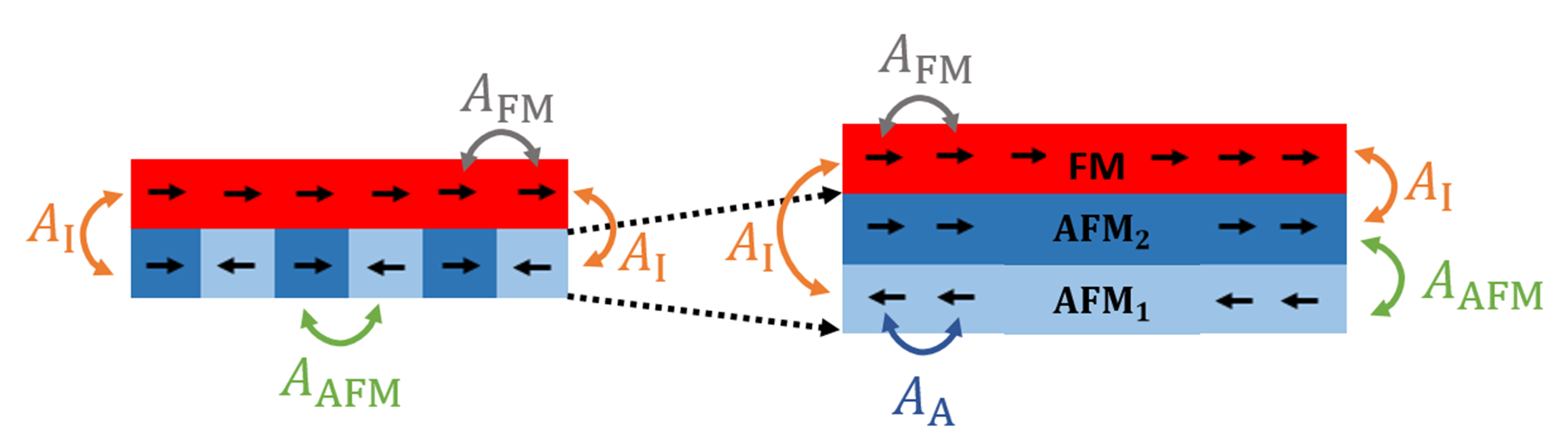}
\caption{In our micromagnetic model (right) of the compensated AFM/FM interface (left), the two sublattices of the atomic AFM layer (with negative exchange stiffness $A_\mathrm{AFM}$) are modelled by 2 micromagnetic layers, AFM$_1$ and AFM$_2$ layers (right, blue) with a positive intralayer exchange stiffness $A_\mathrm{A}$ and coupled by a negative interlayer exchange stiffness $A_\mathrm{AFM}$. The FM layer (red) is directly coupled to both AFM$_1$ and AFM$_2$ by an the interlayer exchange stiffness $A_\mathrm{I}$. To establish the coupling of the FM to the AFM$_1$ layer, either periodic boundary conditions are used or a custom field/energy term is added, see Supplementary Material.}
\label{fig:micro_model} 
\end{figure}

For interfaces with thin FM and AFM layers, single micromagnetic layers can be used. As a micromagnetic cell can only contain one magnetization vector in MuMax$^3$, the coinciding cells of an AFM layer are separated into 2 different layers, denoted AFM$_1$ and AFM$_2$ (see figure \ref{fig:micro_model}). This separation in space has no physical implications. Even though the AFM$_1$ layer (figure \ref{fig:micro_model}) is not directly adjacent to the FM layer, a direct coupling can be established by adding a custom field and energy term (see Supplementary Material). A negative interlayer exchange stiffness $A_\mathrm{AFM}$ will ensure the antiferromagnetic coupling of the sublattices and a positive intralayer exchange stiffness $A_\mathrm{A}$ of the same magnitude will allow for the correct domain wall energy in the AFM (see Supplementary Material).  

\subsection{Limitations of the model}
Due to the micromagnetic approximation of the AFM and its implementation in MuMax$^3$, no high frequency dynamics due to the exchange interaction between the 2 AFM sublattices can be modelled in our simulations. When studying quasistatic configurations resulting from an energy minimisation, precession does not have to be taken into account. This is the case for the problems studied in this paper: exchange bias, spin flop coupling and athermal training.
\\
In case of a thicker antiferromagnet, one can expect that the bulk AFM spins will be pinned along the anisotropy axis and only the AFM spins at the interface region will cant towards the FM. As introduced in the model of Mauri\cite{model_mauri}, a planar domain wall can form, which can be included by adding a third fixed layer\footnote{In this case one has to add an energy term $-J_\mathrm{a} \cos(\phi) - J_\mathrm{a} \cos(\theta)$ in the macrospin model (equation \ref{eq:s_afm}). The parameter $J_\mathrm{a}$ is the surface energy density to form a planar domain wall in the antiferromagnet.} or by subdividing the AFM into several bilayers.

\subsection{Implementation}
\label{sec:impl}
As discussed in section \ref{mm_model}, each AFM layer can be modelled as a pseudo - ferromagnetic layer with thickness $t_\mathrm{AFM}$ and anisotropy constant $K_\mathrm{AFM}$. The interfacial exchange energy $J_\mathrm{I}$ and the energy density $\delta$ linked to the mutual interaction between the 2 AFM spins can be defined in terms of exchange stiffnesses between the FM/AFM and the 2 AFM layers respectively. Using the convention used in MuMax$^3$ for the exchange energy density, one obtains that $A_\mathrm{I} = \frac{J_\mathrm{I}C_z}{2}$ and $A_\mathrm{AFM} = - \frac{\delta t_\mathrm{AFM} C_z}{2}$ with $C_z$ the cell size perpendicular to the interface. Assuming that a micromagnetic system consists of 2 AFM sublattices + 1 FM layer, one can couple the FM layer with the nearest AFM layer (AFM$_2$ in figure \ref{fig:micro_model}) by rescaling the exchange stiffness using the scaling factor $S$ as defined in \cite{MuMax3} and discussed in the Supplementary Material. As in the standard version of MuMax$^3$ only nearest neighbouring cells are taken into account for the evaluation of the exchange energy, one has to couple the other AFM layer, labelled by AFM$_1$ in figure \ref{fig:micro_model}, with the FM layer by using periodic boundary conditions, perpendicular to the FM/AFM interface or by defining a custom field / energy term. The former approach can only be used when the FM consists of only 1 layer while the addition of a custom field / energy term is generally applicable. Demagnetization energy should be turned off in the AFM layers and care should be taken when defining the saturation magnetization $M_\mathrm{AFM}$ in the antiferromagnet. For more information, see Supplementary Material. 
\\ 
It is important to note that one can take into account compensated as well as uncompensated (rotatable as well as pinned) AFM spins in the same micromagnetic simulation using this approach. It suffices to locally decouple one AFM layer of the FM and locally set $\delta = 0$. In this case the sublattice anisotropy constant $K_\mathrm{AFM}$ has to be replaced by the total anisotropy constant of the antiferromagnet. %So one can model a mixed compensated and uncompensated AFM interface using MuMax$^3$.
%and $J_\mathrm{I} \rightarrow \frac{J_\mathrm{I}}{2}$ if one defines the coupling parameter $J_\mathrm{I}$ as defined in the Meiklejohn and Bean model, as in our model each layer acts with a surface energy density $J_\mathrm{I}$ on the FM and  $K_\mathrm{AFM}$ represents the sublattice anisotropy constant.

\subsection{Breakdown of the small canting angle approximation}\label{breakdown_small}
\begin{figure}[htb]
\begin{center}
 \includegraphics[width=0.43\textwidth]{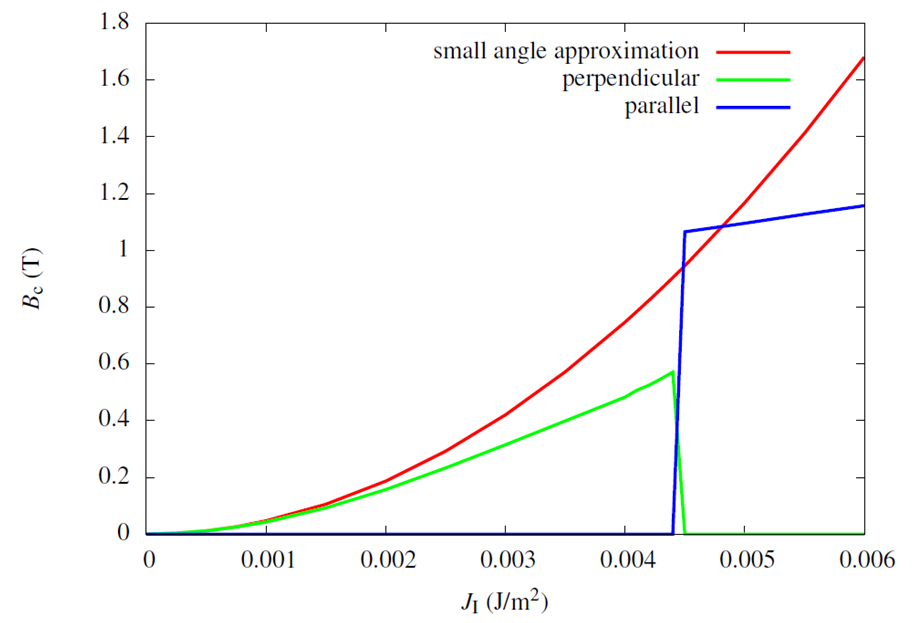}
\caption{Coercivity $B_\mathrm{c}$ as a function of the coupling constant $J_\mathrm{I}$. The small canting angle approximation (red curve) is valid up to $J_\mathrm{I} \approx 2$ mJ/m$^2$. The breakdown of the canted spin flop state happens around $J_\mathrm{I} \approx 4.5$ mJ/m$^2$ and leads to a vanishing coercivity for a hysteresis loop measured perpendicular (green curve) to the N\'eel vector. The blue curve is the coercivity of a hysteresis loop measured parallel to the N\'eel vector.}
\label{fig:breakdown_coerc} 
\end{center}
\end{figure}

\begin{figure}[htb]
\begin{center}
 \includegraphics[width=0.4\textwidth]{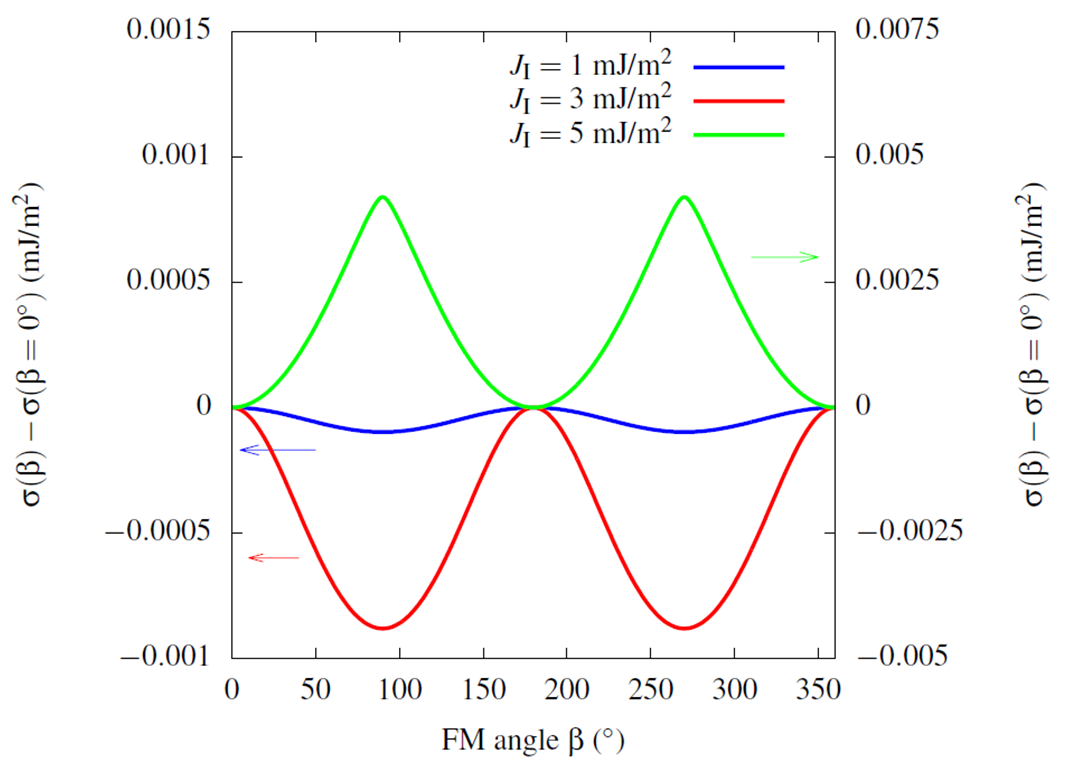}
\caption{Minimal surface energy density curves $\sigma$ as a function of the FM rotation angle $\beta$. A spin flip transition happens for $J_\mathrm{I} \approx 4.5$ mJ/m$^2$ which leads to the fact that the spin flop state is not the global energy minimum anymore. Each curve was shifted by $\sigma (\beta = 0^{\circ})$. The values of the curve for $J_\mathrm{I} = 5$ mJ/m$^2$ correspond to the scale on the right.}
\label{fig:breakdown_energy} 
\end{center}
\end{figure}

To compare the coercivity $B_\mathrm{c}$ derived from the small canting angle approximation (equation \ref{eq:switching}) with the coercivity resulting from the exact energy terms, a simple system was studied with $t_\mathrm{AFM} = t_\mathrm{FM} = 3$ nm, $M_\mathrm{FM} = 1400$ kA/m, $\delta = 1 \times 10^6$ J/m$^3$ and $K_\mathrm{AFM} = 7 \times 10^5$ J/m$^3$. Demagnetization energy was turned off in the FM and no FM anisotropy was considered. To avoid metastable states, each simulation consisted of 2 consecutive hysteresis loops for $H_\mathrm{ext}$ parallel\footnote{In the simulations $H_\mathrm{ext}$ is set at a small angle of 1$^{\circ}$ with the defined directions to introduce a slight asymmetry.} and perpendicular to the uniaxial anisotropy axis of the AFM. The coercivity of the second hysteresis loop together with the small canting angle approximation reported in the previous section are shown in figure \ref{fig:breakdown_coerc}.
\\
One can see that, for these parameters, the small angle approximation of our model is valid up to $J_\mathrm{I} \approx 2$ mJ/m$^2$ where the relative error is around 5 \%. In figure \ref{fig:breakdown_energy}, the minimized total surface energy density (without Zeeman energy) of the system is shown as function of the angle $\beta$ for different coupling constants $J_\mathrm{I}$.
\\
For low $J_\mathrm{I}$, the minima are located at the angles $\beta = \pm \frac{\pi}{2}$, representing the global energy minima. This corresponds to what was discussed in the small canting angle approximation as can be seen in equation \ref{eq:Ku}. Even though the approximation is not valid anymore around $J_\mathrm{I} \approx 2$ mJ/m$^2$, the spin flop state is still the global energy minimum and will lead to a vanishing coercivity for a hysteresis loop, measured parallel to the N\'eel vector, i.e. along one of the AFM easy axes, as can be seen in figure \ref{fig:breakdown_coerc}.

At $J_\mathrm{I} \approx 4.5$ mJ/m$^2$ however, the shape of the energy function changes because the canted spin flop state is not the global energy minimum anymore as a spin flip transition occurs, similarly to the metamagnetic spin flip transition of an antiferromagnet in a strong magnetic field. In this case, the interface coupling $J_\mathrm{I}$ overcomes the intersublattice interaction $\delta$ resulting in a parallel orientation of the 2 AFM layers. Due to the strong magnetocrystalline anisotropy of the AFM, the global energy minima are now given by $\beta =$ 0 and $\pi$ , i.e. parallel to the easy axes of the AFM, and the direction perpendicular to the N\'eel vector becomes a hard axis. This leads to a finite coercivity in the hysteresis loop parallel to the N\'eel vector in figure \ref{fig:breakdown_coerc} and a vanishing coercivity at $J_\mathrm{I} \approx 4.5$ mJ/m$^2$ for the perpendicular hysteresis loop.

\subsection{Interface coupling in LSMO/LFO square nanostructures }
\begin{figure}[htb]
  \centering
  \subfloat[LSMO]{\includegraphics[width=0.12\textwidth]{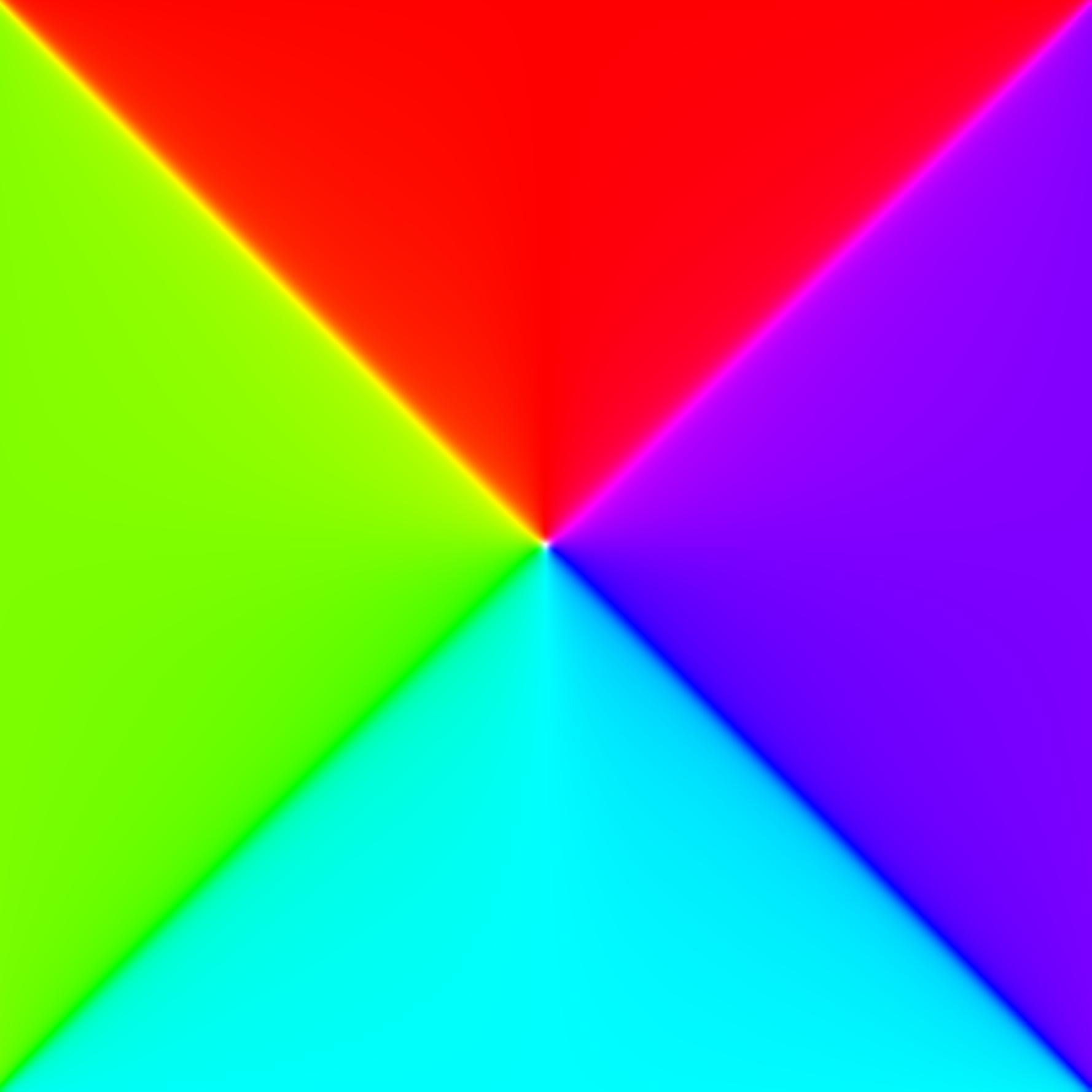}\label{fig:FM_square}}
  \hfill
  \subfloat[LFO layer]{\includegraphics[width=0.12\textwidth]{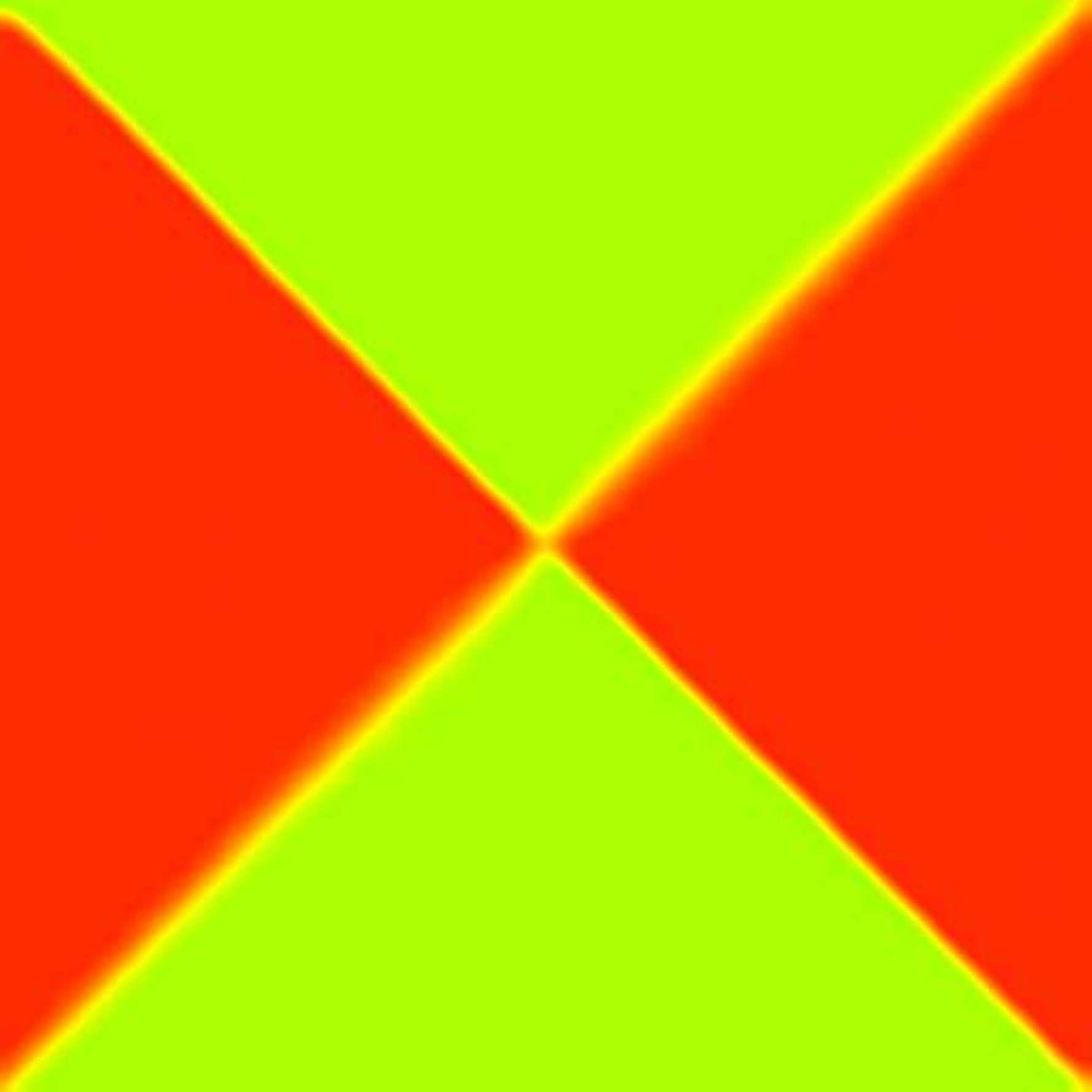}\label{fig:AFM_square}}
  \hfill
  \subfloat[colour scale]{\includegraphics[width=0.12\textwidth]{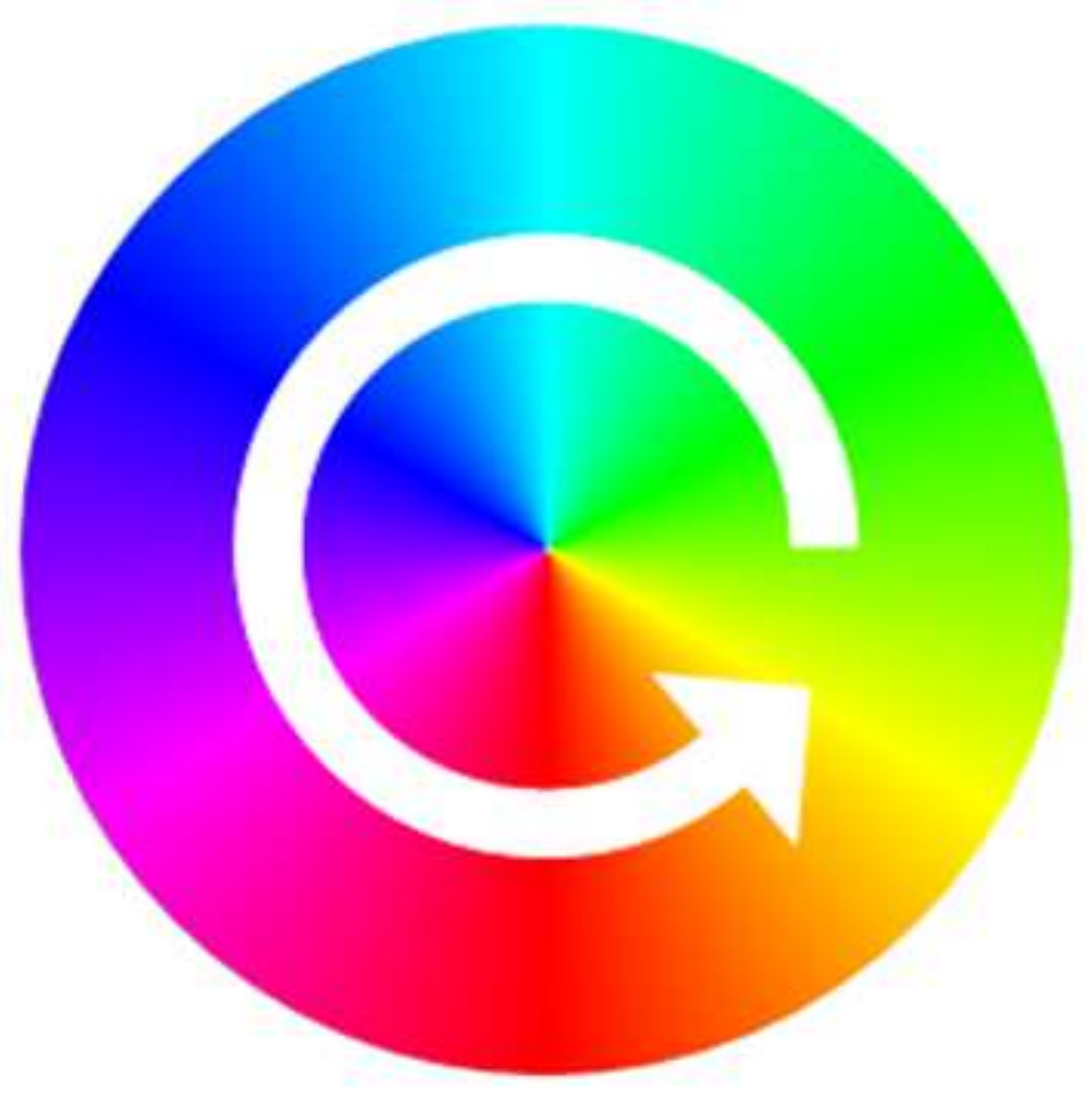}\label{fig:colour_scale}}
  \caption{Spin flop coupling in a LSMO/LFO nanosquare for $B_\mathrm{b} = 0$ mT, initialized from a random state. Figure (a) displays the typical Landau domain structure in the ferromagnetic LSMO. Figure (b) shows the average of the absolute values of the magnetization of the 2 antiferromagnetic LFO layers. In figure (c) the colour scale of the magnetization is displayed. The magnetization vector is tangent to the circle shown.}
\label{fig:square_folven}
\end{figure}

\begin{figure}[htb]
  \centering
  \subfloat[]{\includegraphics[width=0.12\textwidth]{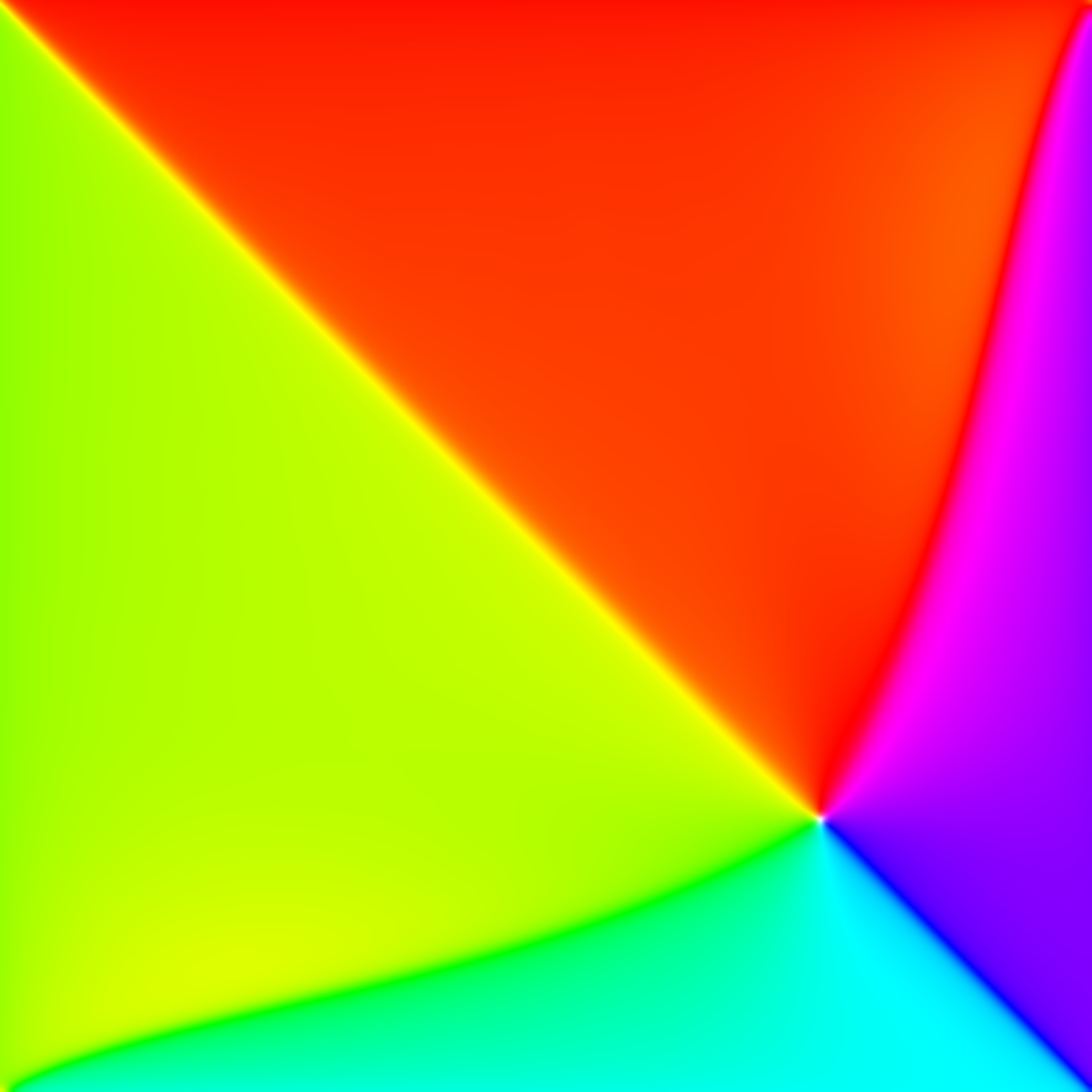}\label{fig:FM_vortex}}
  \hfill
  \subfloat[]{\includegraphics[width=0.12\textwidth]{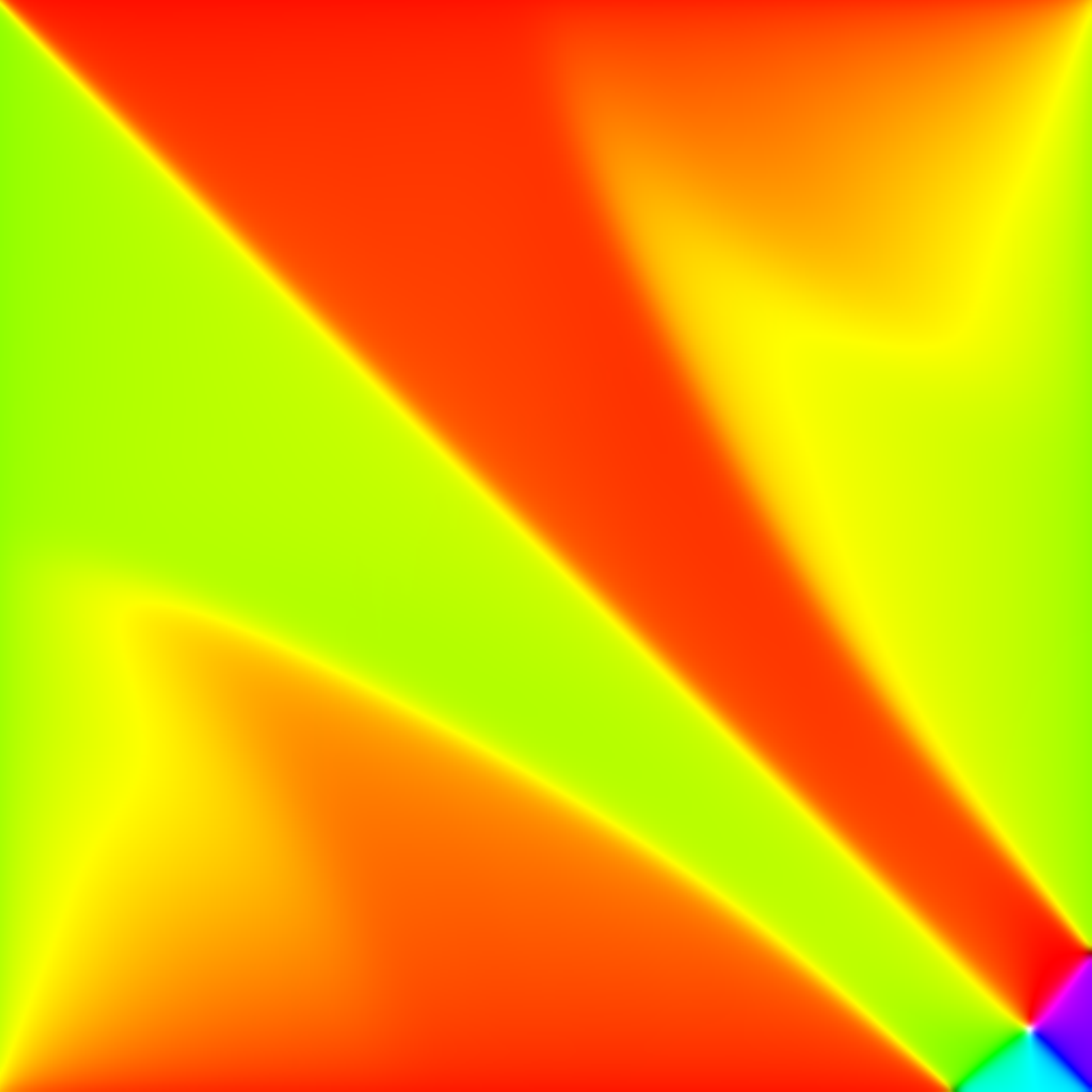}\label{fig:FM_displ}}
    \hfill
  \subfloat[]{\includegraphics[width=0.12\textwidth]{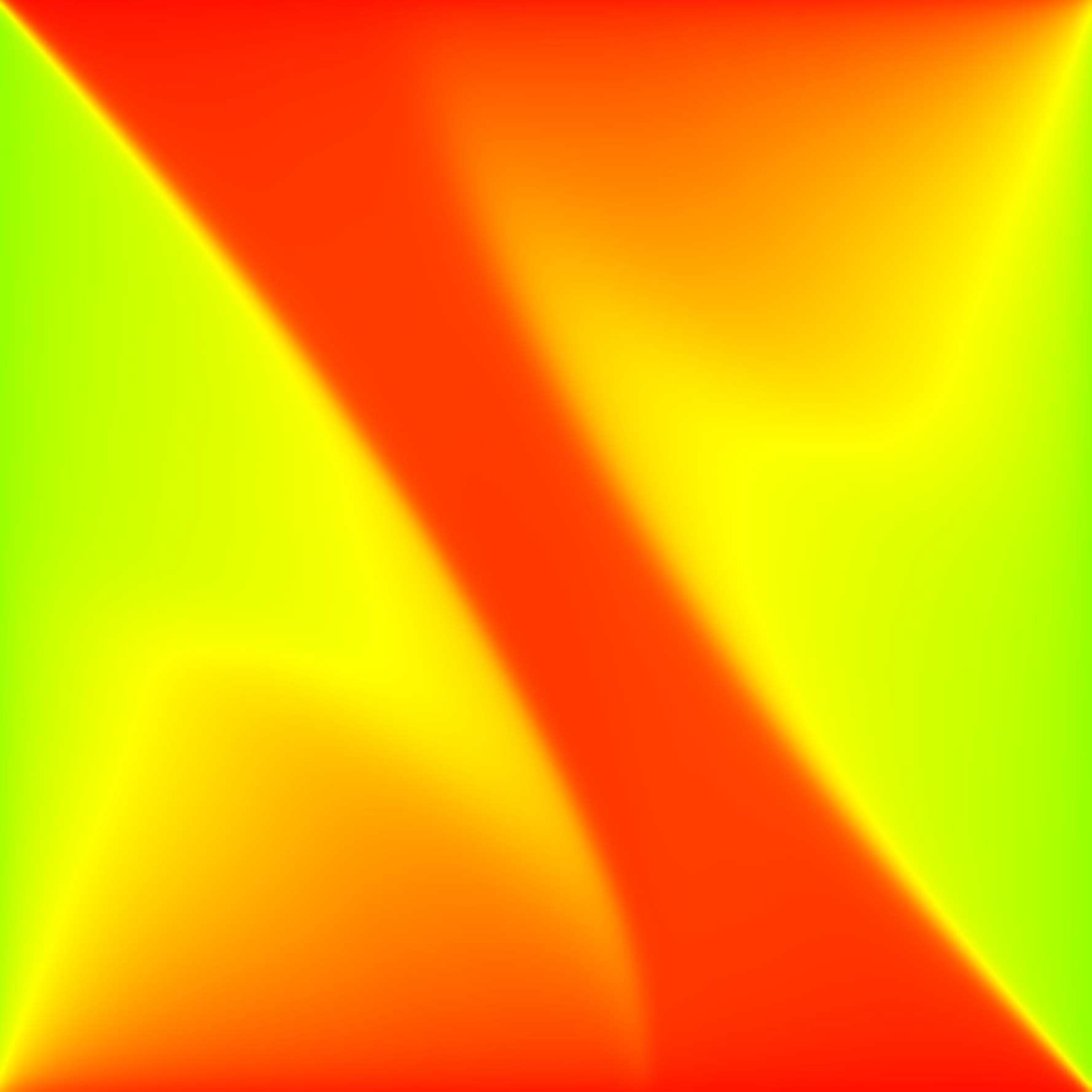}\label{fig:FM_z}}
  \newline
  \subfloat[]{\includegraphics[width=0.12\textwidth]{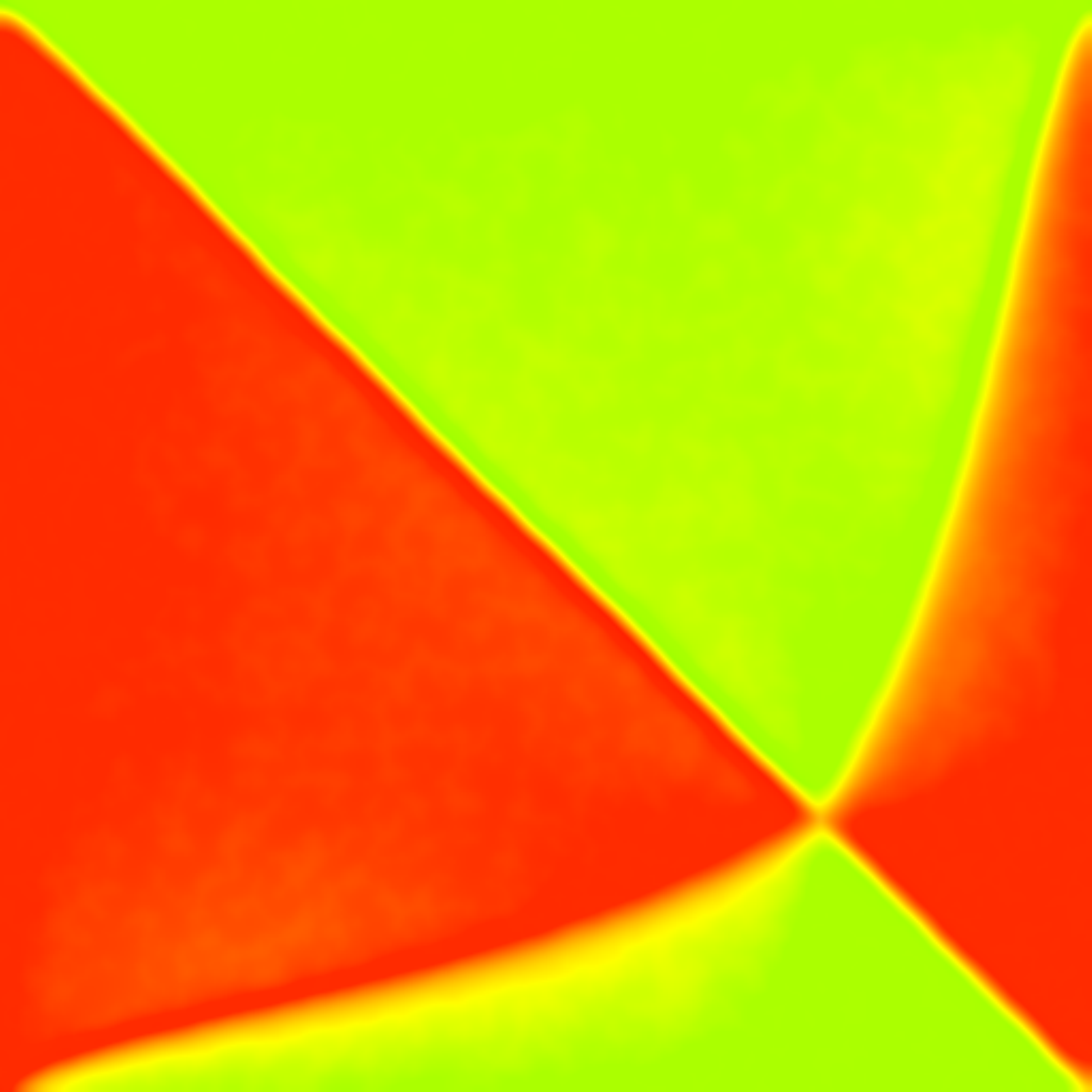}\label{fig:AFM_vortex}}
    \hfill
  \subfloat[]{\includegraphics[width=0.12\textwidth]{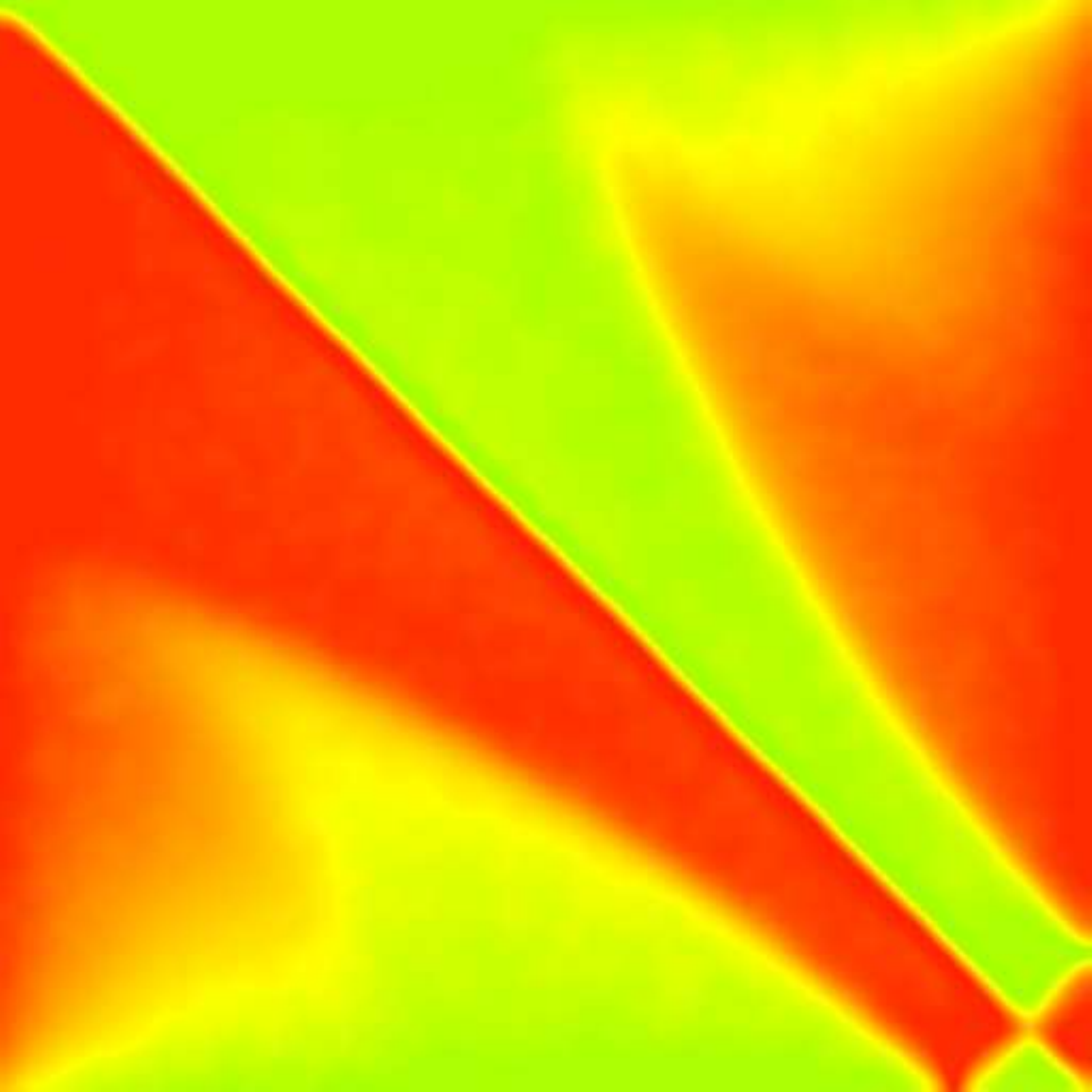}
  \label{fig:AFM_displ}}
  \hfill
  \subfloat[]{\includegraphics[width=0.12\textwidth]{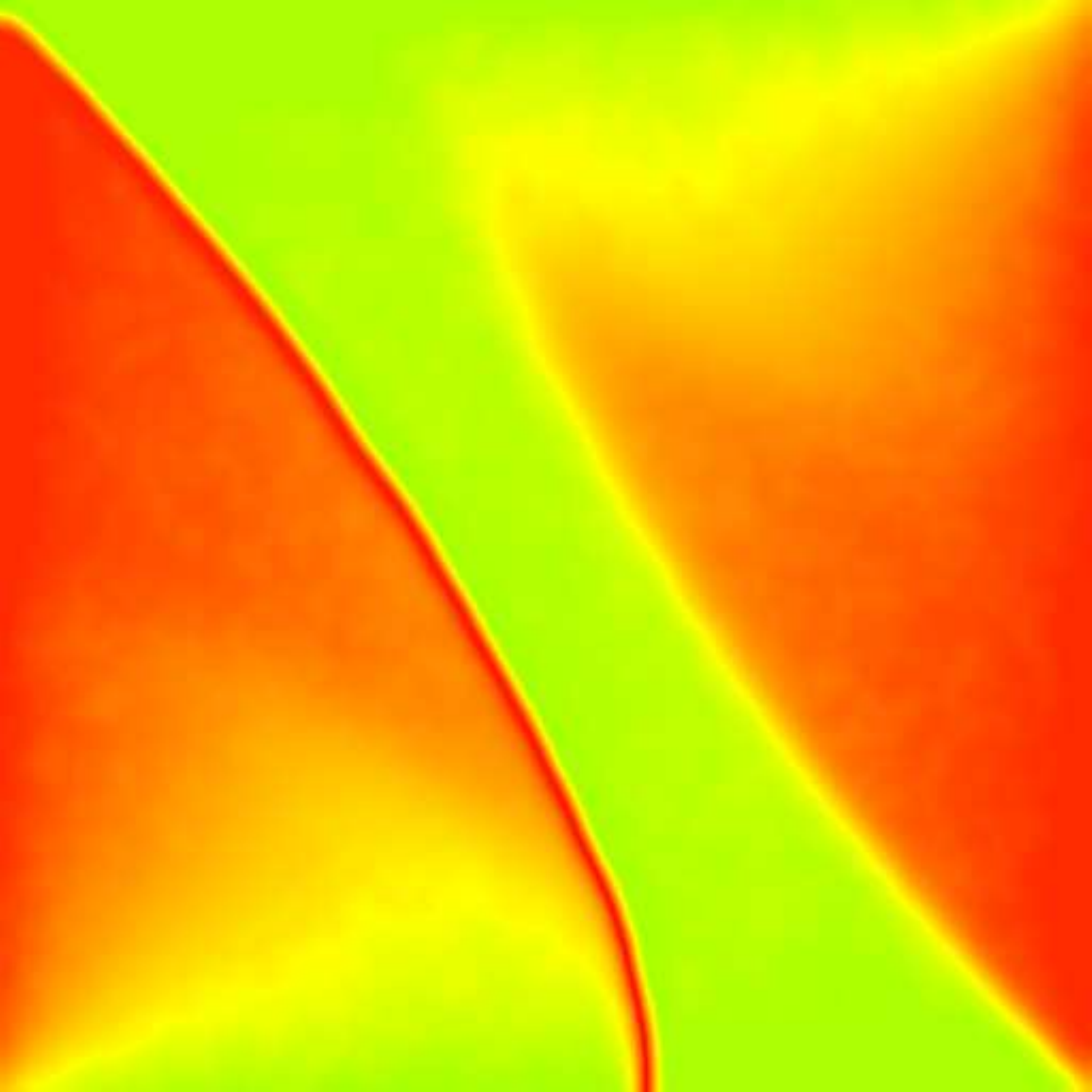}\label{fig:AFM_z}}
  \caption{Magnetic configurations in the LSMO/LFO square for different bias fields $B_\mathrm{b}$. (a) and (d): flux closed Landau state for $B_\mathrm{b} = 6$ mT. (b) and (e): vortex displaced towards the corner for $B_\mathrm{b} = 8$ mT. (c) and (f): z-domain for $B_\mathrm{b} = 10$ mT. The top row represents the FM layer and bottom row represents the corresponding AFM domain structure. Each time the average of the absolute value of the magnetization of the 2 AFM layers has been displayed.}
\label{fig:domains_AFM}
\end{figure}

\begin{figure}[htb]
  \centering
  \subfloat[]{\includegraphics[width=0.12\textwidth]{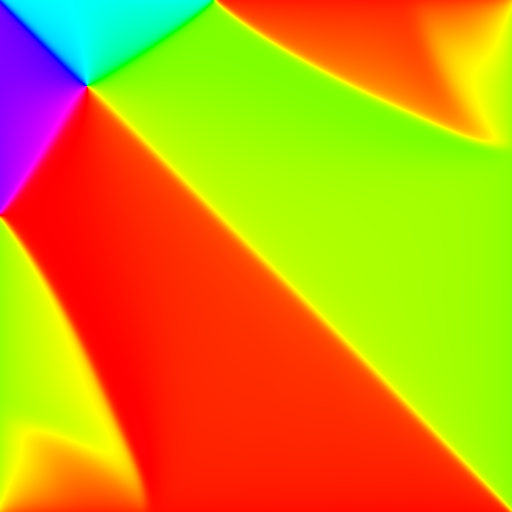}}
  \hfill
  \subfloat[]{\includegraphics[width=0.12\textwidth]{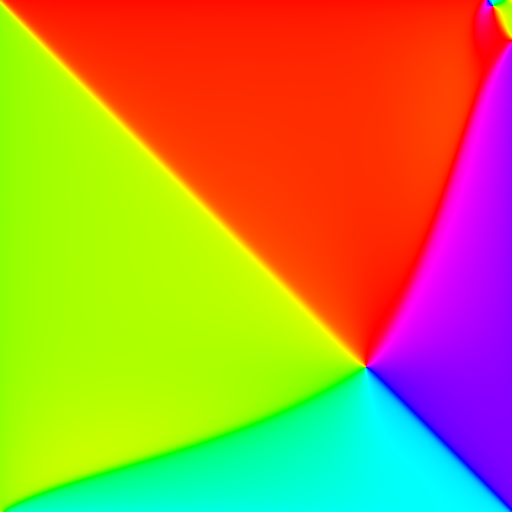}}
    \hfill
  \subfloat[]{\includegraphics[width=0.12\textwidth]{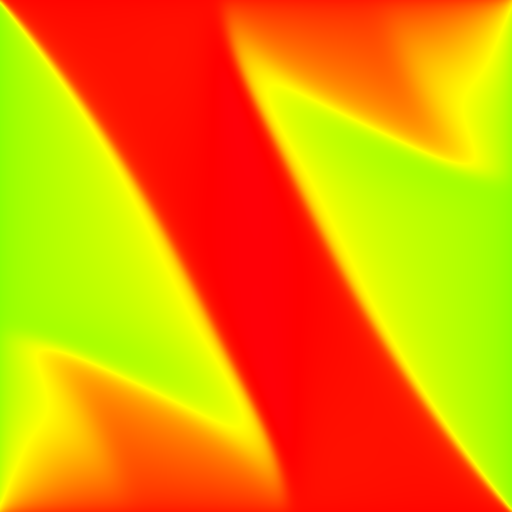}}
  \newline
  \subfloat[]{\includegraphics[width=0.12\textwidth]{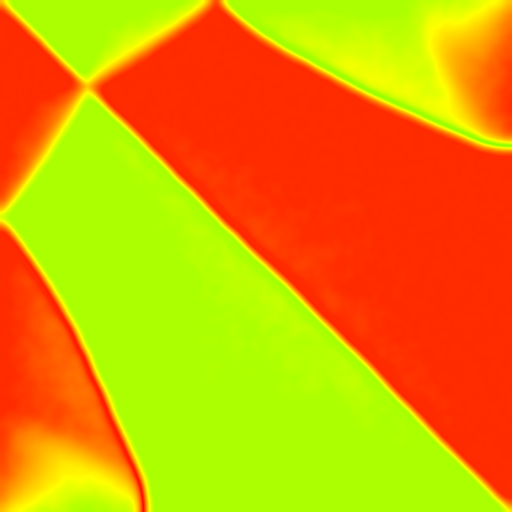}}
    \hfill
  \subfloat[]{\includegraphics[width=0.12\textwidth]{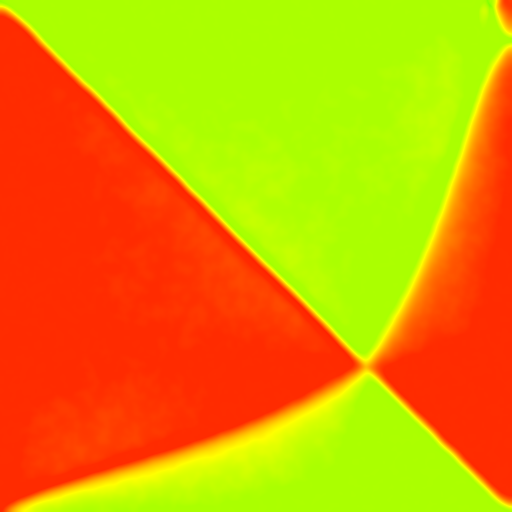}}
  \hfill
  \subfloat[]{\includegraphics[width=0.12\textwidth]{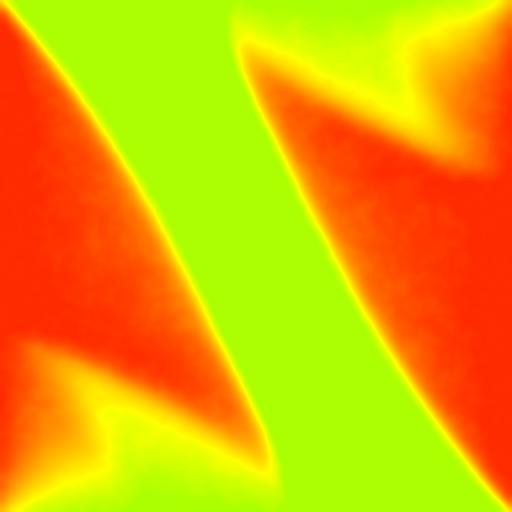}}
  \caption{FM (top) and corresponding AFM (bottom) domain structures for $B_\mathrm{b} = 5$ mT, started from a random configuration in the FM layer initialized with different random seeds. Each time thermal fluctuations were applied for 0.2 $\mu$s before the system was relaxed. For the AFM, each time the average of the absolute value of the magnetization of the 2 AFM layers has been displayed.}
\label{fig:domains_5mT}
\end{figure}

As a first application of our micromagnetic model for spin flop coupling, we reproduce the experimental X-PEEM data of epitaxially grown La$_{0.7}$Sr$_{0.3}$MnO$_3$(35nm) / compensated LaFeO$_3$(3.8nm) square nanostructures as has been reported by Takamura et. al in \cite{vortex_folven}. They noticed that in the bilayer, the N\'eel vector of the antiferromagnetic LFO was oriented perpendicular to the domain structure in the ferromagnetic LSMO square, in accordance with spin flop coupling between the FM and compensated AFM. 
\\
Two types of FM domain structures were observed in the LSMO/LFO squares. The authors argued that these structures result from variations in the local bias field, induced by the DMI interaction. For low bias fields, a Landau structure (with a displaced vortex) is found while higher local bias fields result in a z-type domain. A corresponding perpendicular domain structure was found in the AFM. For a single uncoupled LSMO layer, only the typical Landau domain structure was observed. This implies that variations of the local bias field $B_\mathrm{b}$ can change the domain structure of the LSMO layer and thus also of the AFM.
\\
The ferromagnetic LSMO has biaxial anisotropy with 2 easy axes, denoted by the vectors $\vec{c}_1$ and $\vec{c}_2$, oriented along the $\left<110\right>$ directions, i.e. the sides of the 2x2 $\mu$m$^2$ square. The anisotropy energy density is given by $\epsilon_\mathrm{K} = K_\mathrm{c} (\vec{c}_1.\vec{m})^2(\vec{c}_2.\vec{m})^2$ with $\vec{c}_1$ pointing along the $x$-axis and $\vec{c}_2$ along the $y$-axis of the simulation box. The bilayer was divided into 512 x 512 x (2 AFM + 1 FM) cells with a lateral cell size of 3.9 nm and thickness\footnote{For a rescaling of the AFM parameters to achieve the correct total energy, please see Supplementary Material. } of 35 nm. For the ferromagnetic LSMO we used typical parameters: $M_\mathrm{FM} = 400$ kA/m, $A_\mathrm{FM} = 1.8$ pJ/m and $K_{\mathrm{c},\mathrm{FM}} = 1.6$ kJ/m$^3$. 
\\
Experiments\cite{vortex_folven,biax_folven} show that 60 \% of the AFM domains in the LFO layer have their biaxial easy axes along the $\left<110\right>$ directions and the remaining 40 \% have their easy axes along the $\left<100\right>$ directions. Therefore, the anisotropy axes were distributed accordingly for the AFM layer in our simulations by using a Voronoi tesselation\footnote{It is clear that the direction of the anisotropy axes, etc. in an AFM grain should be set equal in the 2 AFM layers.}. We used the parameters: $\delta = 1.25 \times 10^5$ J/m$^3$, $K_\mathrm{c,AFM} = 1.5$ kJ/m$^3$ and $J_\mathrm{I} = 0.17$ mJ/m$^2$ for the interaction between the AFM and FM. 
\\
Starting from a random FM state and semi-random\footnote{30\% of the AFM grains were initialised along the $x$-axis, 30\% along the $y$-axis, 20\% along one diagonal of the square and 20\% along the other diagonal. The sublattices were initialised each time antiparallel to each other. } AFM state, the system was relaxed towards equilibrium after being subjected to thermal fluctuations (70 K during 0.2 $\mu$s). 
The FM and the average of the absolute value of the magnetisation of the 2 AFM layers (at remanence) are shown in figure \ref{fig:square_folven}. 
In this case, the AFM orients itself perpendicular to the FM as the ferromagnetic domain structure is determined by the demagnetization energy. A similar effect was also observed in NiO/Fe and CoO/Fe nanodisks.\cite{vortex_spin_flop}
\\
In the experimental data shown in figure 3 of \cite{vortex_folven} however, also z-type of domains in the FM and corresponding perpendicularly coupled domains in the AFM were found. By comparing the data to a micromagnetic simulation of only the FM layer\footnote{See figure 4c and 4d in \cite{vortex_folven}}, the authors state that these domain structures appear for bias fields above a threshold of approximately 9 mT and correspond to a vortex, displaced towards the corner of a square. To test this hypothesis, a hysteresis loop of the FM/compensated AFM square was simulated with an external field, representing the bias field $B_\mathrm{b}$, applied along one of the diagonals. Starting from remanence, as shown in figure \ref{fig:square_folven}, the flux closed vortex state is stable for $B_\mathrm{b} < 8$ mT. For a bias field of 8 mT, the vortex gets displaced towards the corner of the square. For $B_\mathrm{b} > 8$ mT, z-domains\footnote{When the LSMO is not coupled to the LFO, no bias field can be present and thus only Landau states are formed in the ferromagnet.} are formed in the FM and a corresponding perpendicular domain structure is found in the AFM layers\footnote{The domain wall in the AFM is induced by the initial Landau state.}. These 3 magnetic configurations are shown in figure \ref{fig:domains_AFM}.
\\
When returning from saturation however, the FM does not return towards the vortex state for $B_\mathrm{b} < 8$ mT. The presumption thus arises that these z-domains can also be present in lower bias fields, as opposed to what was claimed by the authors in \cite{vortex_folven}. When initializing the FM from completely random states, only Landau domain structures were found for $B_\mathrm{b} = 0$ mT. Relaxing different random states in a bias field $B_\mathrm{b} = 5$ mT, z-domains as well as domains with a displaced vortex were formed. The results are shown in figure \ref{fig:domains_5mT}. Note that two variations of the z-domain can be found due to the symmetry of the system, in correspondence to the experimental data. When $B_\mathrm{b} \geq 7.5$ mT only z-domains were formed. This shows that the creation of z-domains are indeed intimately linked to the presence of a large bias field, although they can be stable for lower bias fields as well. 

\subsection{Athermal training effects}
\label{S:4}
\subsubsection{Hoffmann training}
As a second example we will take a look at the training effect in a compensated AFM, as was proposed by Hoffmann\cite{Hoffmann} and is nowadays generally accepted.\cite{cit_Hoffman,cit_Hoffman2} While there can be no training effects or exchange bias in a perfectly compensated antiferromagnet with uniaxial anisotropy, Hoffmann argued that this is not the case when the AFM has a fourfold (or higher) magnetocrystalline anisotropy. He argued that, after field cooling, the sublattices of the AFM can be in a non collinear state and only relax towards an antiparallel configuration after the first reversal of the FM. This produces an athermal training effect and exchange bias in the first hysteresis loop ($n = 1$). Also the asymmetry between the first drop of the hysteresis loop and its reversal towards positive saturation is a prominent feature of athermal training.

\subsubsection{Phase diagram of Hoffmann training}
\begin{figure}[htb]
\begin{center}
 \includegraphics[width=0.4\textwidth]{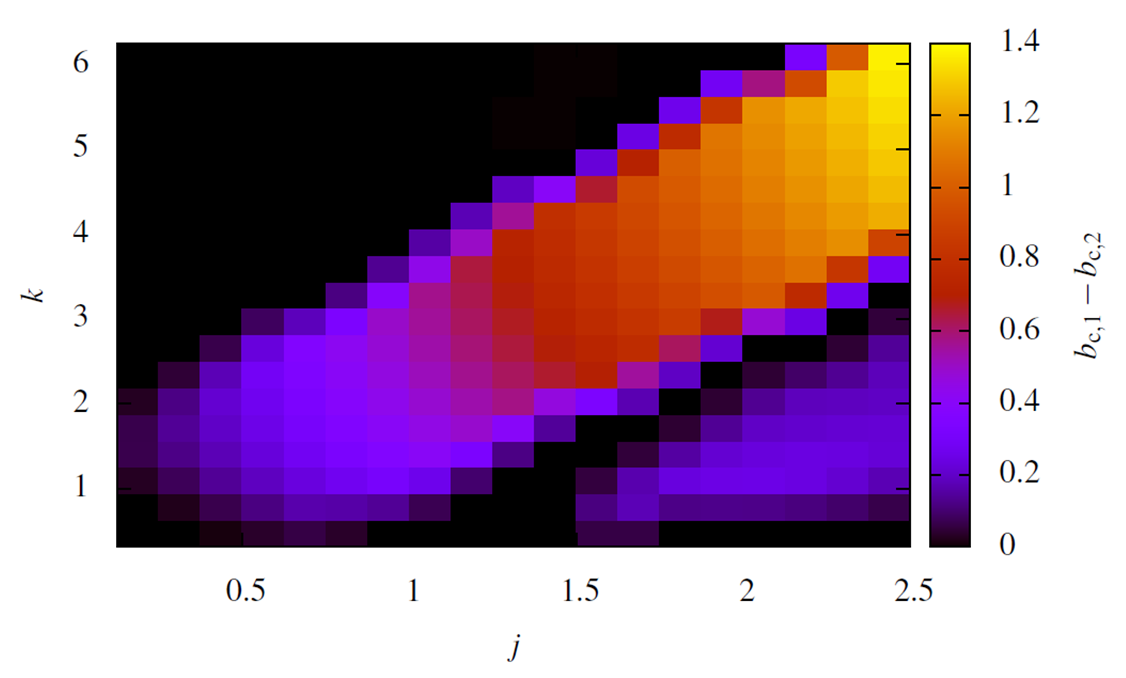}
\caption{Phase diagram of Hoffmann training for a uniform biaxial antiferromagnet, as a function of the reduced anisotropy constant $k$ and the reduced exchange coupling $j$. The colour scale (on the right) represents the difference in coercivity $b_\mathrm{c}$ (in reduced units) between the first and second hysteresis loop. }
\label{fig:parameter_scan_Hoffmann} 
\end{center}
\end{figure}

\begin{figure}[htb]
  \centering
  \subfloat[$k = 0.625$]{\includegraphics[width=0.11\textwidth]{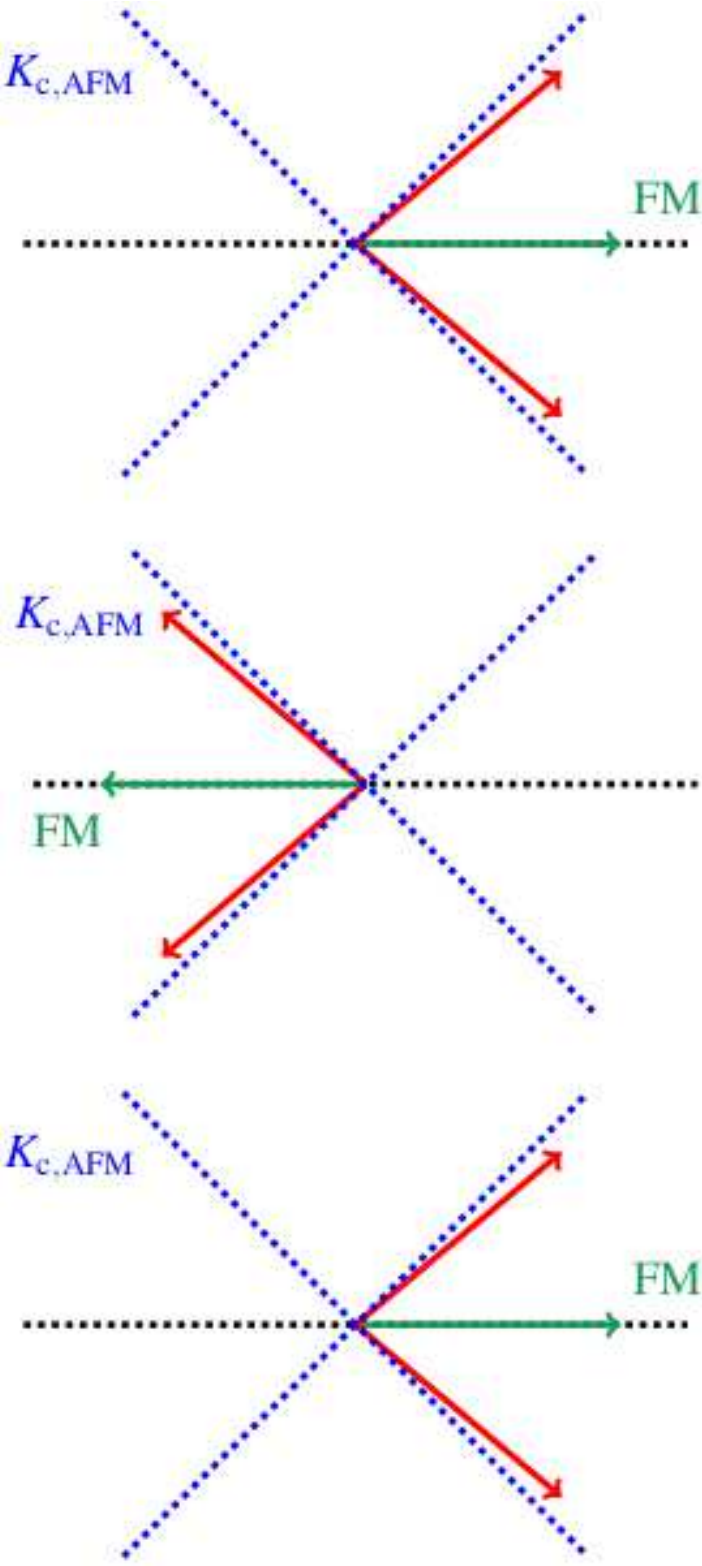}\label{fig:k0_625}}
  \hfill
  \subfloat[$k = 1$]{\includegraphics[width=0.11\textwidth]{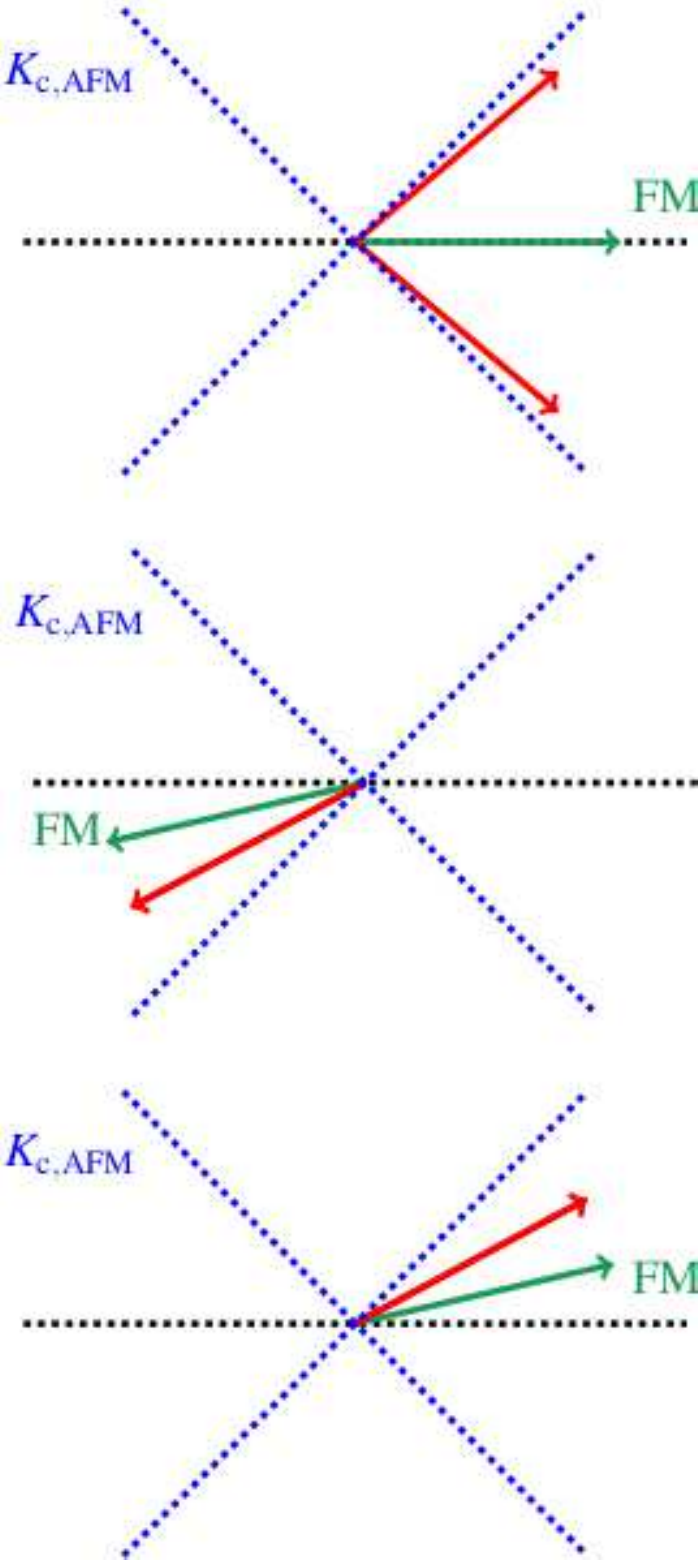}\label{fig:k1}}
  \hfill
  \subfloat[$k=4$]{\includegraphics[width=0.11\textwidth]{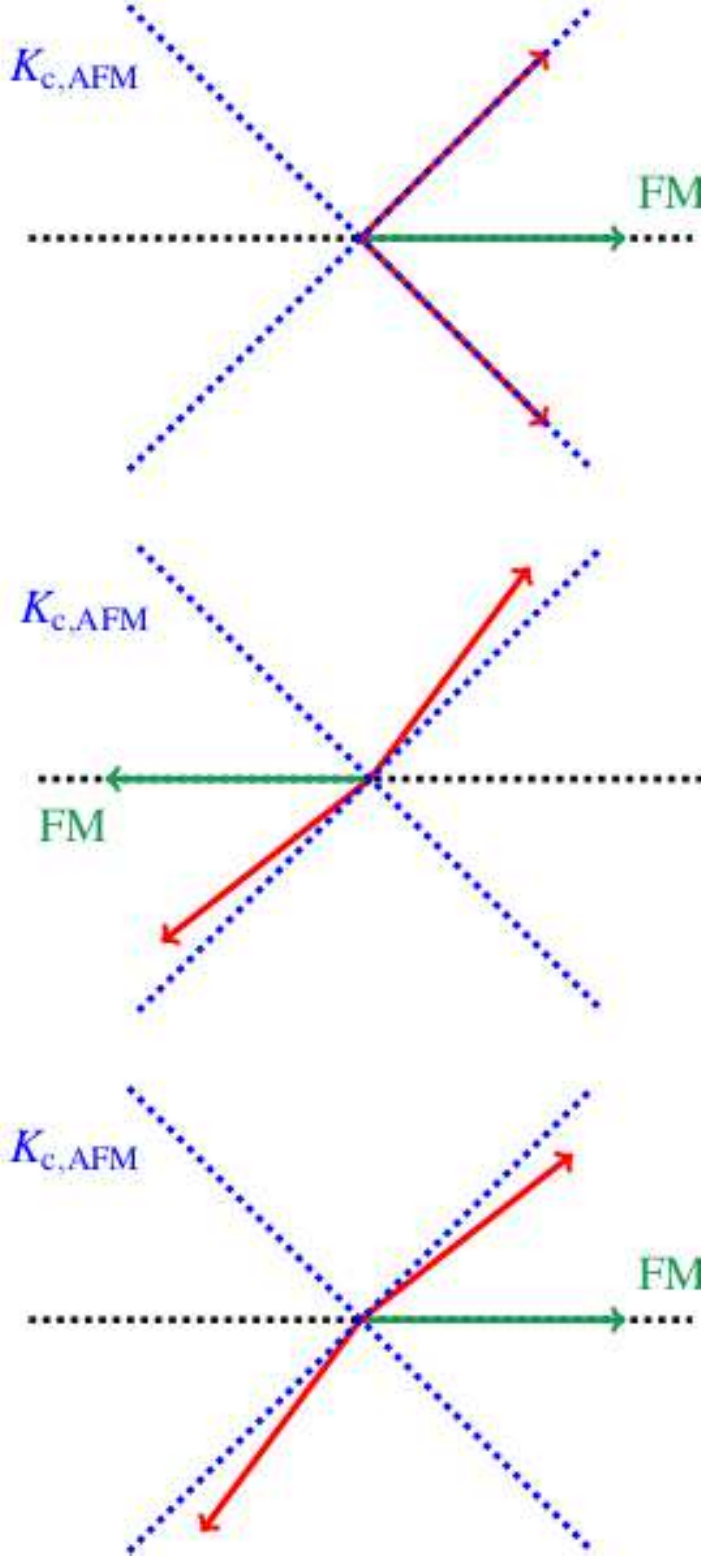}
  \label{fig:k4}}
    \hfill
  \subfloat[$k=6$]{\includegraphics[width=0.11\textwidth]{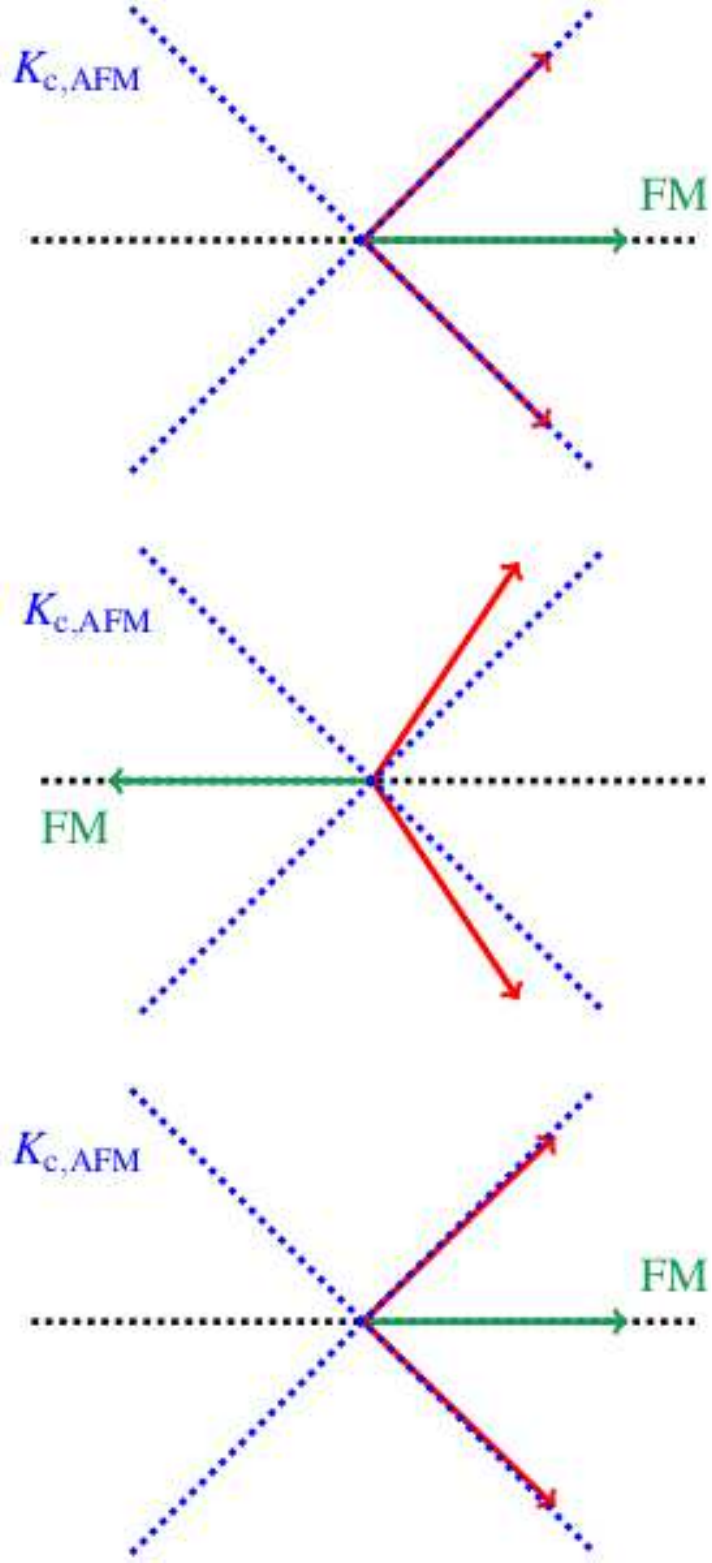}\label{fig:k6_25}}
  \caption{Each column represents a different region in the phase diagram (figure \ref{fig:parameter_scan_Hoffmann}) for $j=1.75$. Top row: relaxed field cooled state, middle row: state at negative saturation, bottom row: state at positive saturation again. The green arrow represents the FM and the red arrows represent the 2 AFM layers.}
\label{fig:phase_arrow}
\end{figure}

As the training effect in the model of Hoffmann results from a non collinear state of the 2 AFM sublattices induced by field cooling, we can expect that training will only happen within a certain parameter range. As an example, we will investigate this effect by considering a uniform biaxial compensated antiferromagnet whose easy axes make an angle of 45$^{\circ}$ degrees with respect to the field cooling direction. To simulate field cooling, the AFM layers were initialized in a spin flop state, i.e. in the (1,1,0) and (1,-1,0) directions along their easy axes. Afterwards the system was relaxed while the FM was saturated in the field cooling direction. A cell size of 2 nm in the in-plane direction and a cell thickness of 10 nm was chosen in a direction perpendicular to the interface, i.e. in this case $t_\mathrm{FM} = t_\mathrm{AFM} =$ 10 nm. For the FM typical parameters of Py were used: $M_\mathrm{FM}$ = 800 kA/m, $A_\mathrm{FM}$ = 13 pJ/m and magnetocrystalline anisotropy was not considered. For the antiferromagnet we used $\delta = 8 \times 10^4$ J/m$^3$. The biaxial anisotropy constant $K_\mathrm{c,AFM}$ and the coupling parameter $J_\mathrm{I}$ were varied and the coercivity and exchange bias field were determined for 2 consecutive hysteresis loops along the field cooling direction.\footnote{$H_\mathrm{ext}$ is set at a small angle of 1$^{\circ}$ with the field cooling direction to introduce a slight asymmetry.} The Zeeman energy was not taken into account for the AFM. 
\\
The result of this parameter scan is shown in figure \ref{fig:parameter_scan_Hoffmann} where the difference in coercivity between the first and second hysteresis loop is displayed. Reduced dimensionless units were used, i.e. 
\begin{align*}
j = \frac{J_\mathrm{I}}{\delta t_\mathrm{AFM}} \quad \quad k = \frac{K_\mathrm{c,AFM} t_\mathrm{AFM}}{\delta t_\mathrm{AFM}}\quad \quad b_\mathrm{c} = \frac{\mu_0 M_\mathrm{FM} t_\mathrm{FM} H_\mathrm{c}}{\delta t_\mathrm{AFM}}
\end{align*}
for the interface coupling, anisotropy and coercivity, respectively. 
The different reversal mechanisms, present in the phase diagram, are illustrated in figure \ref{fig:phase_arrow} for a coupling constant $j=1.75$. 

\begin{figure}[htb]
\begin{center}
 \includegraphics[width=0.35\textwidth]{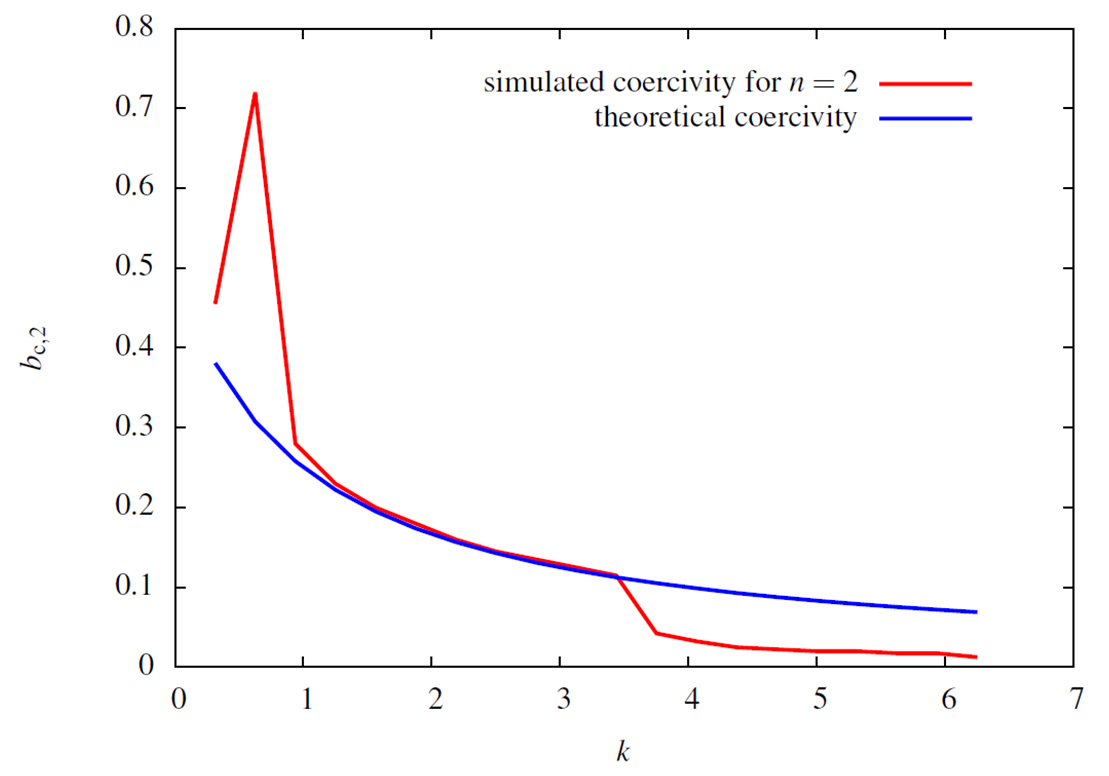}
\caption{Comparison of the approximate theoretical coercivity $b_\mathrm{c}$ (for $j=1$) as follows from the Stoner Wohlfarth model and the simulated coercivity for $n=2$. The theoretical curve matches with the simulated one in case the 2 AFM layers stay in an antiparallel position for $n=2$, thus from $k \approx 0.75$ to $k \approx 3.5$. }
\label{fig:spinflopbiaxiaal} 
\end{center}
\end{figure}

\begin{figure}[htb]
\begin{center}
 \includegraphics[width=0.35\textwidth]{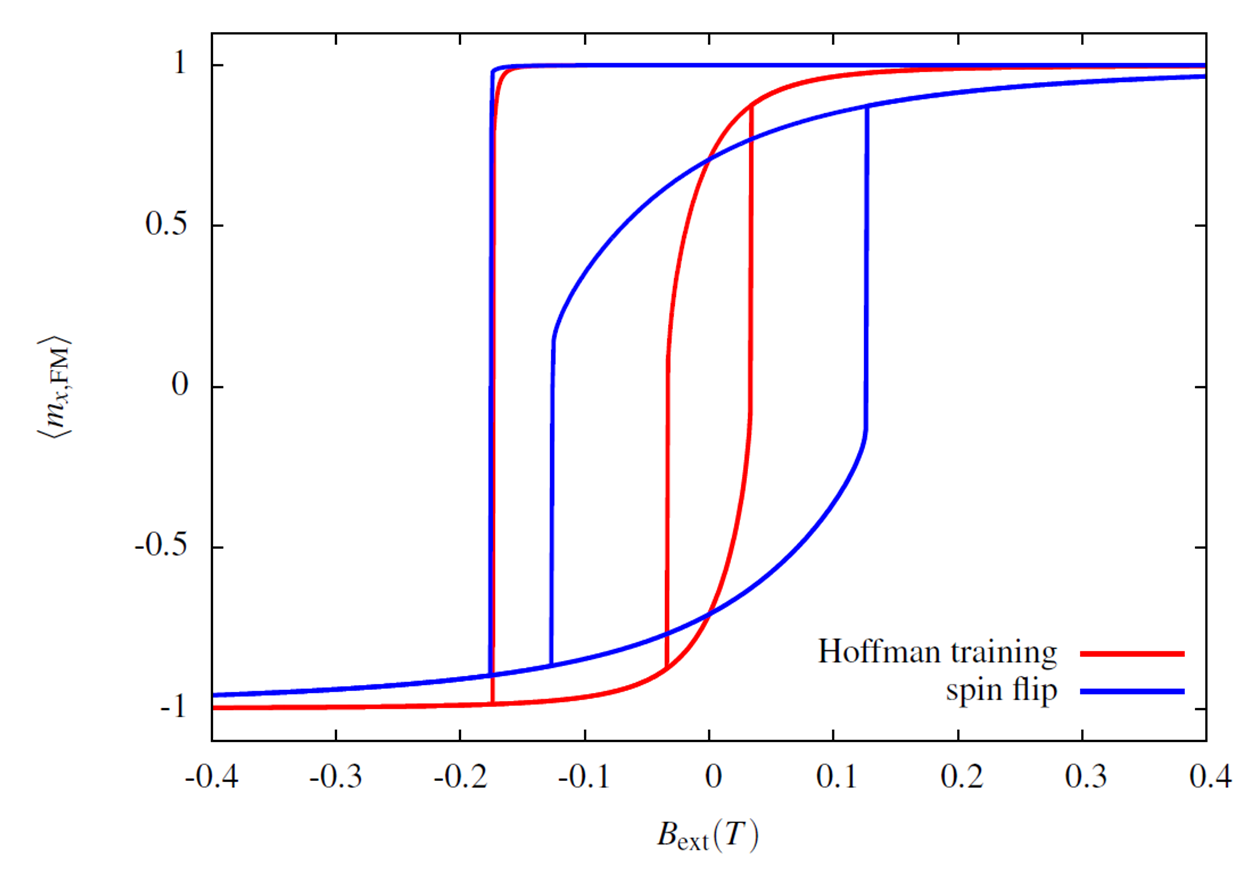}
\caption{FM hysteresis loop, measured along the field cooling direction in the region of Hoffmann training ($j = 1.5$, $k=2.5$) and in the region of spin flip transition ($j = 2.25$, $k = 1.25$). In both cases the asymmetry and training effect are clearly present.}
\label{fig:hysteresis_Hoffmann} 
\end{center}
\end{figure}

\begin{figure}[htb]
\begin{center}
 \includegraphics[width=0.4\textwidth]{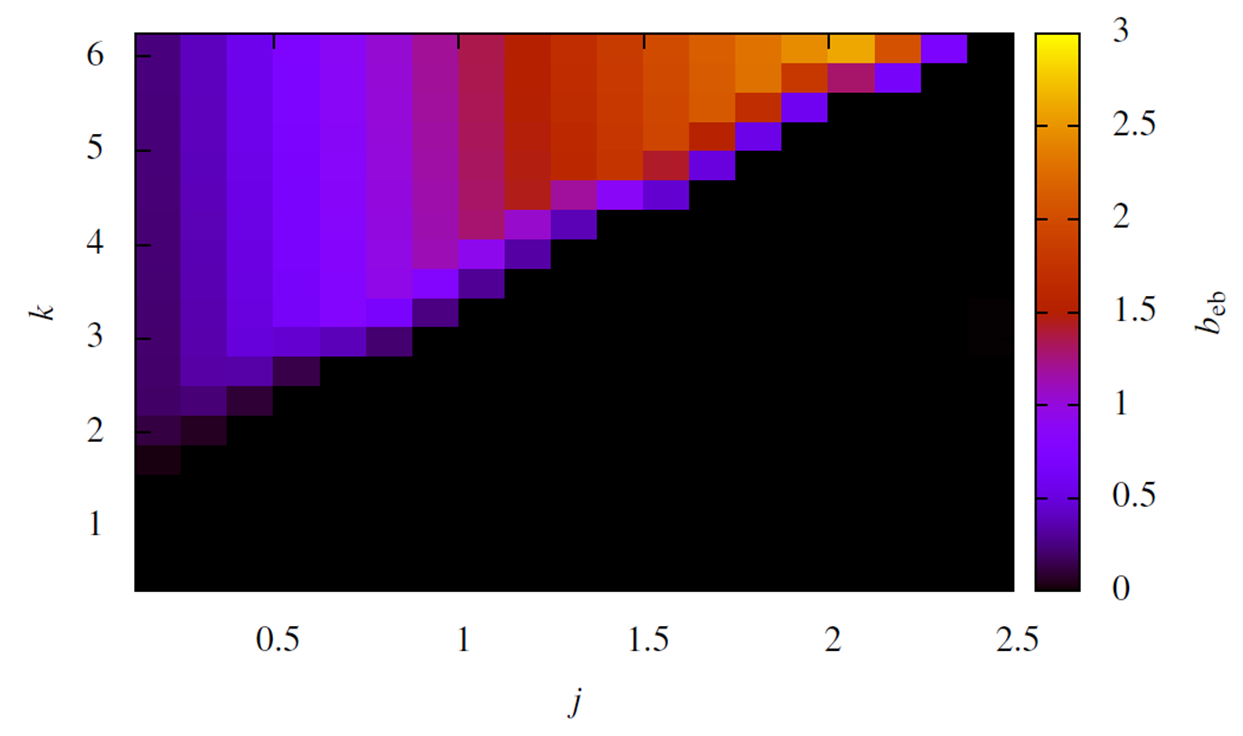}
\caption{Regions in the phase diagram (for $n=2$) which lead to permanent exchange bias due to pinned AFM spins. The figure shows the reduced bias field $b_\mathrm{eb}$ for $n=2$, as a function of the reduced anisotropy constant $k$ and the reduced exchange coupling $j$. The colour scale is to the right.}
\label{fig:eb_Hoffmann} 
\end{center}
\end{figure}

\begin{figure}[htb]
\begin{center}
 \includegraphics[width=0.335\textwidth]{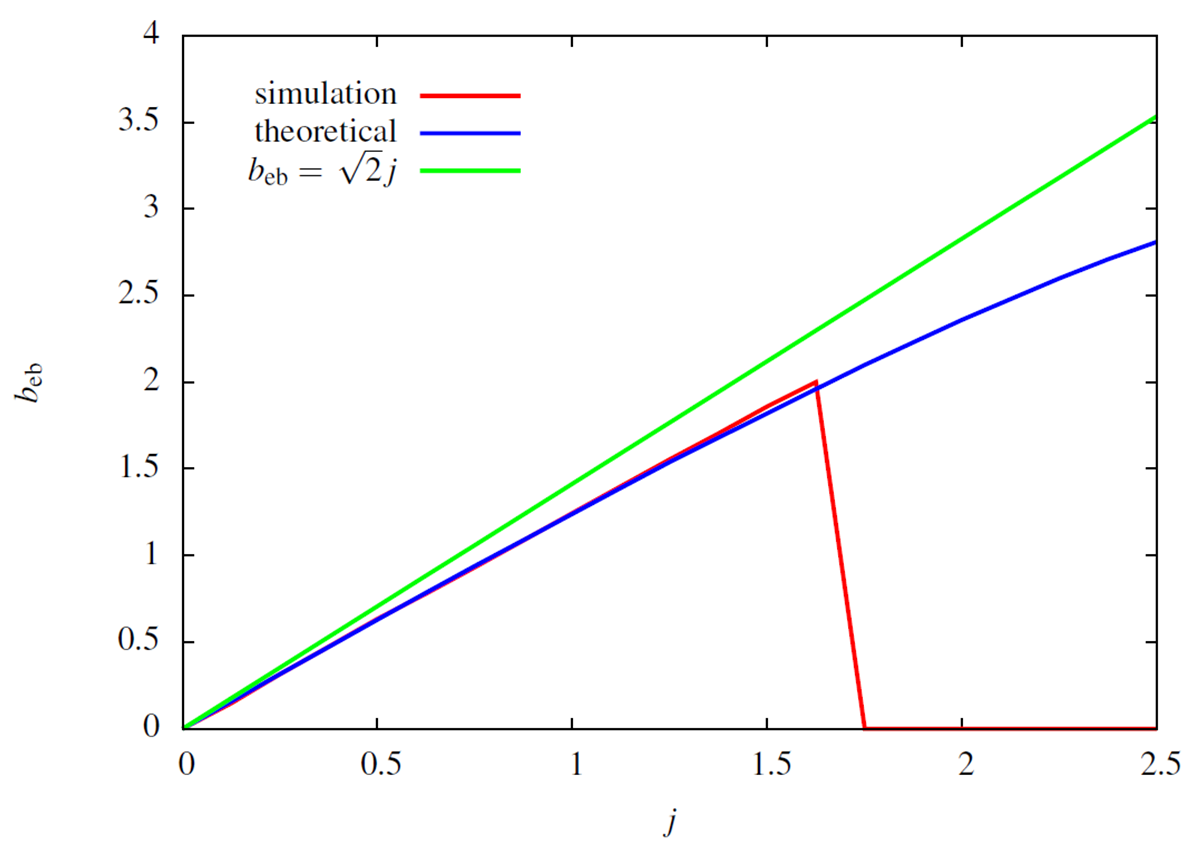}
\caption{Comparison of the approximated theoretical bias field and the values obtained from our simulations ($n=2$) for $k = 5$. Each time, absolute values for the exchange bias field are shown.}
\label{fig:eb_comp} 
\end{center}
\end{figure}

\begin{figure}[htb]
\begin{center}
 \includegraphics[width=0.39\textwidth]{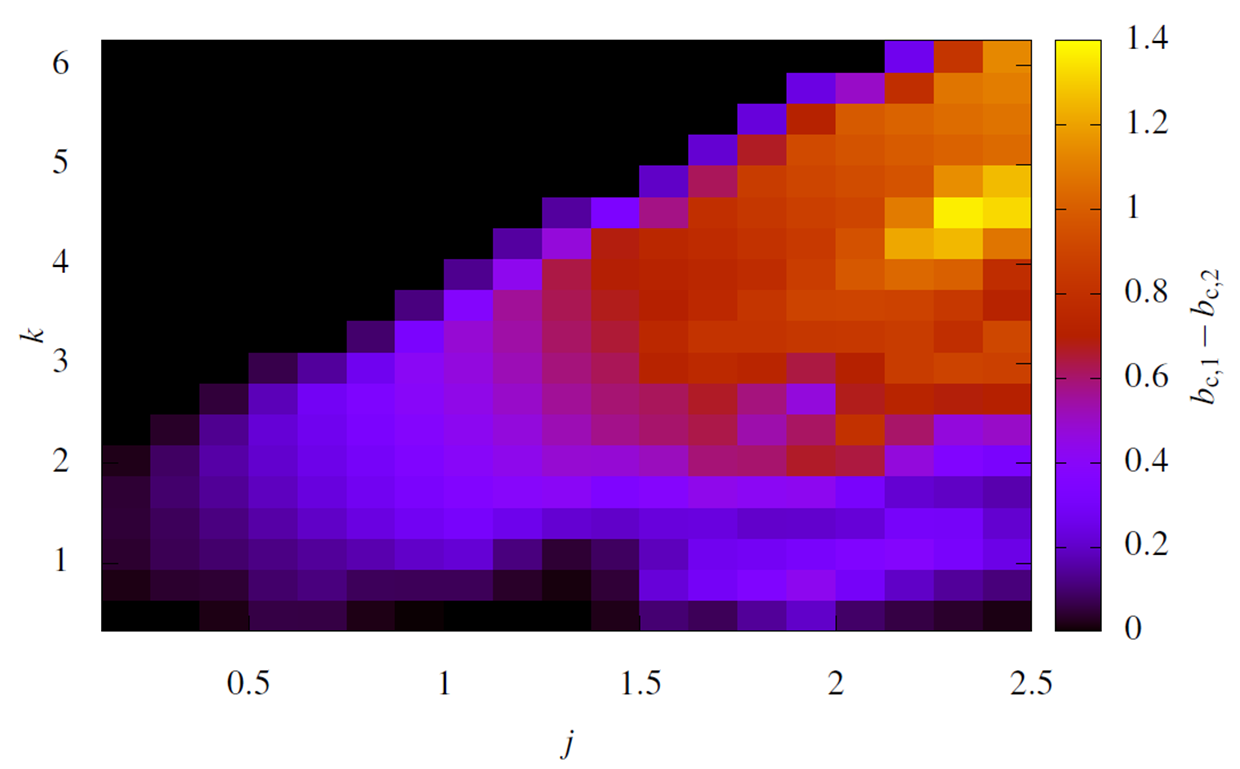}
\caption{Phase diagram of Hoffman training for a polycrystalline biaxial antiferromagnet, as a function of the reduced anisotropy constant $k$ and the reduced exchange coupling $j$. The colour scale (on the right) represents the difference in coercivity $b_c$ (in reduced units) between the first and second hysteresis loop. The same parameters were used as in the uniform case, shown in figure \ref{fig:parameter_scan_Hoffmann}.}
\label{fig:Bcpolybiax} 
\end{center}
\end{figure}

\begin{figure}[htb]
\begin{center}
 \includegraphics[width=0.39\textwidth]{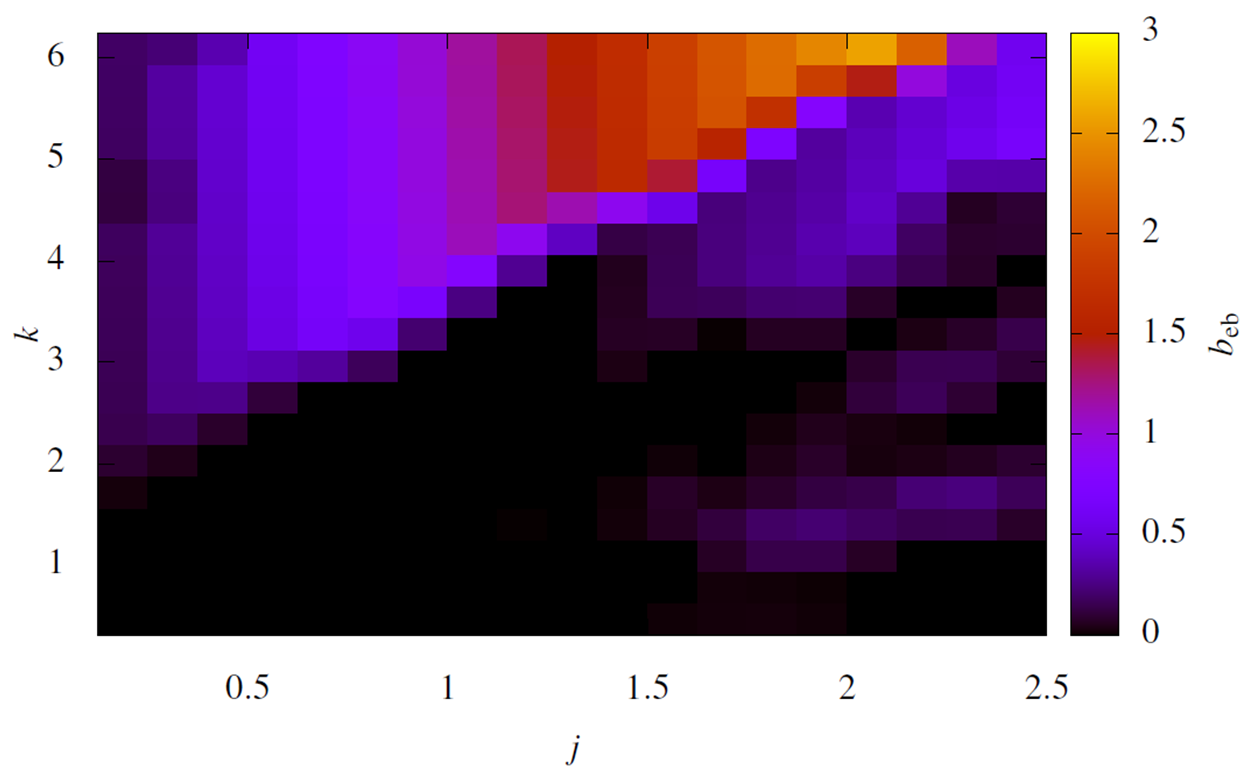}
\caption{Exchange bias field $b_\mathrm{eb}$ for $n=2$ in a polycrystalline AFM for grains with biaxial anisotropy, as a function of the reduced anisotropy constant $k$ and the reduced exchange coupling $j$. The same parameters were used as in the uniform case shown in figure \ref{fig:parameter_scan_Hoffmann}. The colour scale is to the right.}
\label{fig:Bebpolybiax}
\end{center}
\end{figure}

\begin{figure}[htb]
\centering
\includegraphics[width=0.3\textwidth]{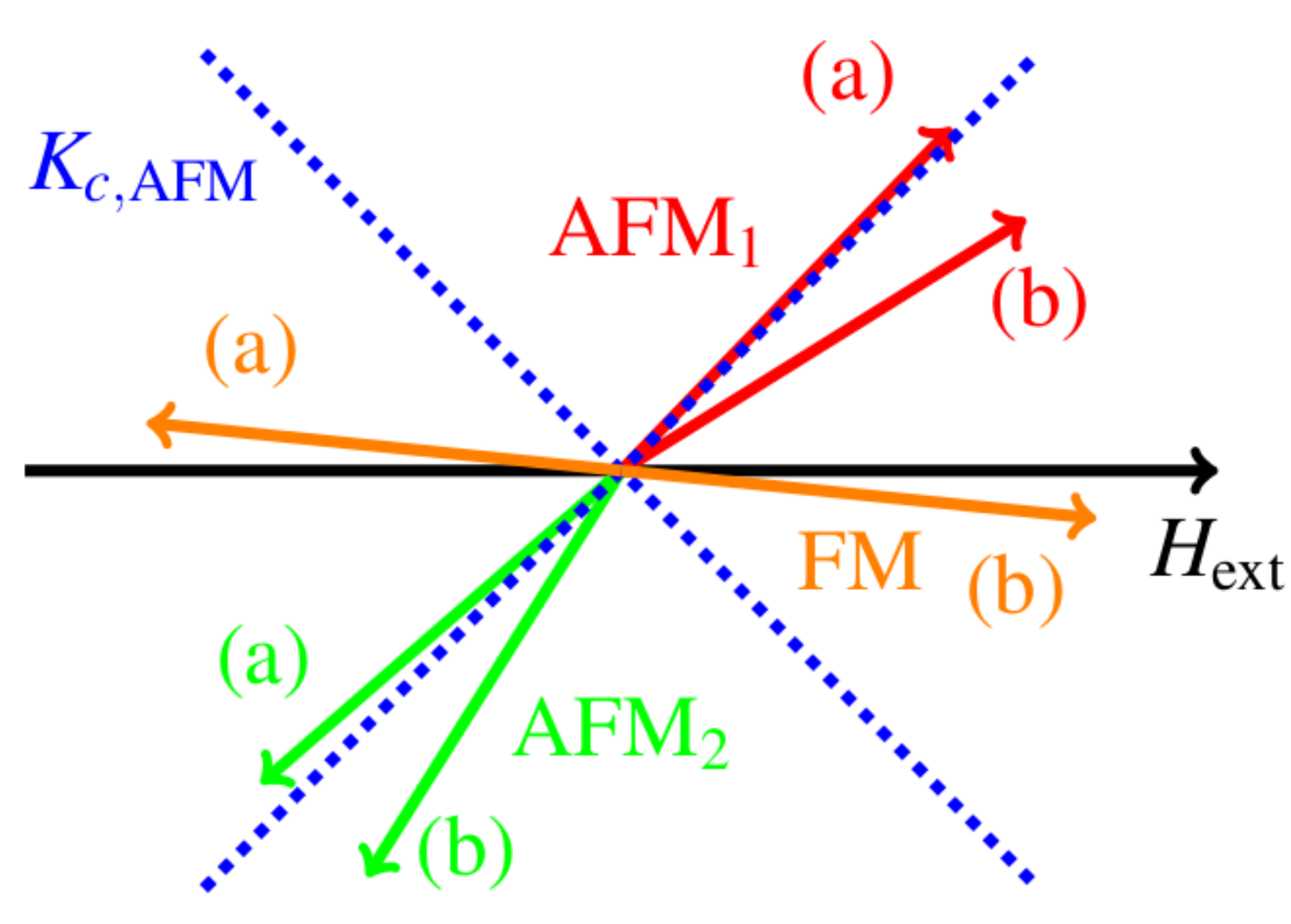}
\caption{Average behaviour of the 2 AFM layers ($j=2$ and $k=4.1$) at negative and positive saturation, labelled by (a) and (b) respectively. Note that the vectors shown only indicate the orientation of the average magnetizations of the AFM$_1$ and AFM$_2$ layers and not the magnitude. The angle the FM makes with respect to the external field is a bit exaggerated. The biaxial anisotropy axes (blue) are shown as a comparison.}
\label{fig:grain_poly} 
\end{figure}

One can clearly distinguish two regions displaying training effects in the phase diagram. The middle region is the real Hoffmann training (figure \ref{fig:k4}) as explained in the previous subsection: the AFM layers change from a non collinear to an almost antiparallel arrangement after the first drop in the hysteresis loop. After this irreversible transition, the AFM layers will induce spin flop coupling (see Supplementary Material) in the FM layer, but with an easy axis rotated over 45$^{\circ}$ with respect to the field cooling direction. According to the Stoner Wohlfarth model, they will also contribute to the coercivity in the second hysteresis loop, but its value will be reduced by a factor 2. The coercivity $b_\mathrm{c,2}$ of the second hysteresis loop (after training) for $j = 1$ as a function of the reduced anisotropy constant $k$ can be seen in figure \ref{fig:spinflopbiaxiaal}. Hoffmann training is present from $k \approx 0.75$ to $k \approx 3.5$. The rise in coercivity for $k < 0.75$ is due to the fact that both AFM layers are approximately in a 90$^{\circ}$ canted position and switch together with the FM. For this configuration (figure \ref{fig:k0_625}), an increasing anisotropy constant leads to an increasing coercivity, in contrast to the case of spin flop coupling.
\\
The training effect in the lower right part (high coupling and low anisotropy) of the phase diagram (figure \ref{fig:parameter_scan_Hoffmann}) is the result of a spin flip transition. After reaching negative saturation in the first hysteresis loop, the two AFM layers stay parallel to each other in further hysteresis loops. This mechanism is displayed in figure \ref{fig:k1}. A similar effect was already seen when considering the breakdown of the canted spin flop state for an AFM with uniaxial anisotropy (section \ref{breakdown_small}). In this case the AFM layers switch irreversibly with the FM for $n > 1$, which leads to a higher coercivity. This is not the case for Hoffmann training. Both regions in the phase diagram however display the asymmetry in the first hysteresis loop which is typical for athermal training, as can be seen in figure \ref{fig:hysteresis_Hoffmann}. 
\\
Athermal training also leads to exchange bias in the first hysteresis loop due to a change in coercivity. For high anisotropy constants $k$ and low coupling constants $j$, i.e. in the upper left part of the phase diagram, one can in fact obtain exchange bias for $n \geq 1$ as the AFM spins are pinned in a spin flop state along the field cooling direction as is displayed in figure \ref{fig:k6_25}. The phase diagram for exchange bias in the case $n \geq 1$ is shown in figure \ref{fig:eb_Hoffmann}. Using the small angle approximation in an AFM with biaxial anisotropy, one can calculate (see Supplementary Material) the bias field $B_\mathrm{eb}$ for high anisotropy constants. In figure \ref{fig:eb_comp} one can see that there is a good agreement between this model and the values obtained from our simulations. Also the bias field for 2 fixed uncompensated spins, which make an angle of 45$^{\circ}$ with the field cooling direction, is shown as comparison (green line in figure \ref{fig:eb_comp}). We remark that in an experiment, one cannot distinguish between a frozen uncompensated AFM spin or 2 compensated biaxial AFM spins, frozen into a 90$^\circ$ canted state, as both have the same net effect on the FM layer. 
\\
If one assumes a negative coupling constant $J_\mathrm{I}$ and takes into account the Zeeman energy for the AFM, one can also obtain positive bias fields, similar to the model that was proposed by Kiwi\cite{Kiwi}. For low cooling fields $H_\mathrm{fc}$, the pinned compensated AFM spins will be antiparallel to the cooling field direction. For stronger cooling fields however, the AFM spins can make an irreversible transition of 90$^{\circ}$ due to the Zeeman energy and produce a net magnetic moment parallel to $H_\mathrm{fc}$. As hysteresis loops are often recorded for $H_\mathrm{ext} << H_\mathrm{fc}$, these spins stay pinned in the spin flop configuration due to the strong magnetocrystalline anisotropy and low coupling to the FM, even during reversal of the ferromagnet.
\\ 
Additionally, a system with a polycrystalline biaxial AFM layer was simulated in the same parameter range as for the uniform case. For each parameter set, the same grains with the same anisotropy axes were used. The AFM intergrain interaction was not taken into account. The difference in coercivity between the first and second hysteresis loop is shown in figure \ref{fig:Bcpolybiax}. Apart from small fluctuations, the global shape of the phase diagram is very similar to figure \ref{fig:parameter_scan_Hoffmann}. The distinct separation between the region of Hoffmann training and the spin flip arrangement is less clear due to the random distribution of the anisotropy axes. 
The bias field in the second hysteresis loop ($n=2$) is shown in figure \ref{fig:Bebpolybiax} and is also similar to figure \ref{fig:eb_Hoffmann}. In some regions there is still a non vanishing bias field for $n\geq 2$. 
The bias field in the region $j=2$, $k=4$ originates from the fact that the absolute value of the average $x$-component of the AFM\footnote{We define $\left< m_{x,\mathrm{AFM}} \right>$  as the quantity $ \left< m_{x,\mathrm{AFM}} \right>  =  \left< m_{x,\mathrm{AFM}_1}\right> +  \left<m_{x,\mathrm{AFM}_2} \right>$.} is asymmetric at positive and negative saturation. In the uniform case we find $\left< m_{x,\mathrm{AFM}} \right> \approx 0.24$ and $\left< m_{x,\mathrm{AFM}} \right> \approx- 0.24$ for positive saturation and negative saturation, respectively. In the polycrystalline case (figure \ref{fig:grain_poly}) however, we retrieve that $\left< m_{x,\mathrm{AFM}} \right>  \approx 0.03$ for negative saturation, but  $ \left< m_{x,\mathrm{AFM}} \right>  \approx 0.41$ for positive saturation. As $ \left< m_{x,\mathrm{AFM}} \right> > 0$ for positive as well as negative saturation, a net average torque is applied on the FM layer and thus exchange bias is produced. The average $y$-components are symmetric at positive and negative saturation, but with opposite signs. This asymmetry can be induced by the finite number of AFM grains in our simulation box and a finite distribution of biaxial anisotropy axes which thus leads to a net preferred direction.
\\
In the region $j=2$, $k=1.6$ a spin flip transition of the 2 AFM layers occurs, analogous to the uniform case. The moments of the 2 AFM layers inside an AFM grain will be parallel after the first reversal ($n=1$) towards negative saturation. The total average $y$-component of the antiferromagnet are $\left< m_{y,\mathrm{AFM}} \right>  \approx -0.12$ and $\left< m_{y,\mathrm{AFM}} \right>  \approx -0.34$ at positive and negative saturation, respectively. The spin flip state together with the effect of the limited size produces also here a small exchange bias.
\\
One can conclude that the average behaviour of a polycrystalline AFM is analogous to the case where the two biaxial anisotropy axes are set symmetrical along the field cooling direction.

\subsubsection{Reproducing athermal training in an IrMn/CoFe bilayer}
As another illustration of training effects in a compensated AFM, we will reproduce the experimental hysteresis loop of an IrMn(15nm)/CoFe(10nm) bilayer, measured at 15 K, as was reported by Fulara et al. \cite{training_IrMn} (see inset figure 1b). Although IrMn has uniaxial anisotropy at high temperatures, experimental evidence\cite{IrMn_biax_proof,training_IrMn} shows that the AFM undergoes a phase transition at 50K and develops a biaxial anisotropy during field cooling. The observed training is attributed to Hoffmann training as it does not follow the power law for thermal training. 

\begin{figure}[htb]
\begin{center}
 \includegraphics[width=0.35\textwidth]{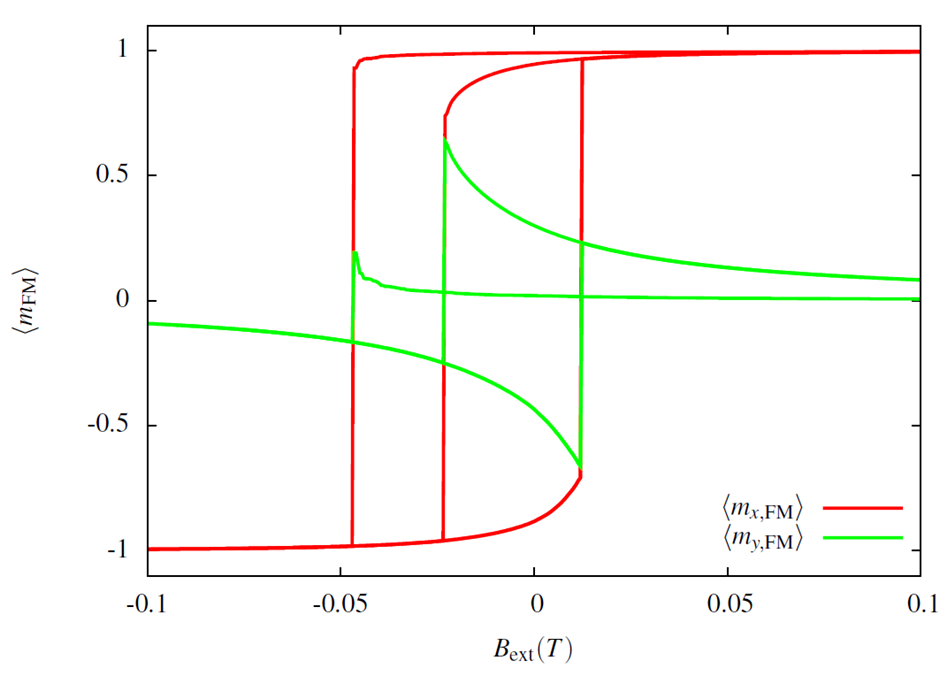}
\caption{Simulated first and second hysteresis loop of an IrMn(15nm)/CoFe(10nm) bilayer. The average $x$ and $y$-components of the FM magnetization are shown. The external field was applied along the $x$-axis, which corresponds to the field cooling direction.} 
\label{fig:CoFe} 
\end{center}
\end{figure}

\begin{figure}[h]
\centering
\includegraphics[width=0.25\textwidth]{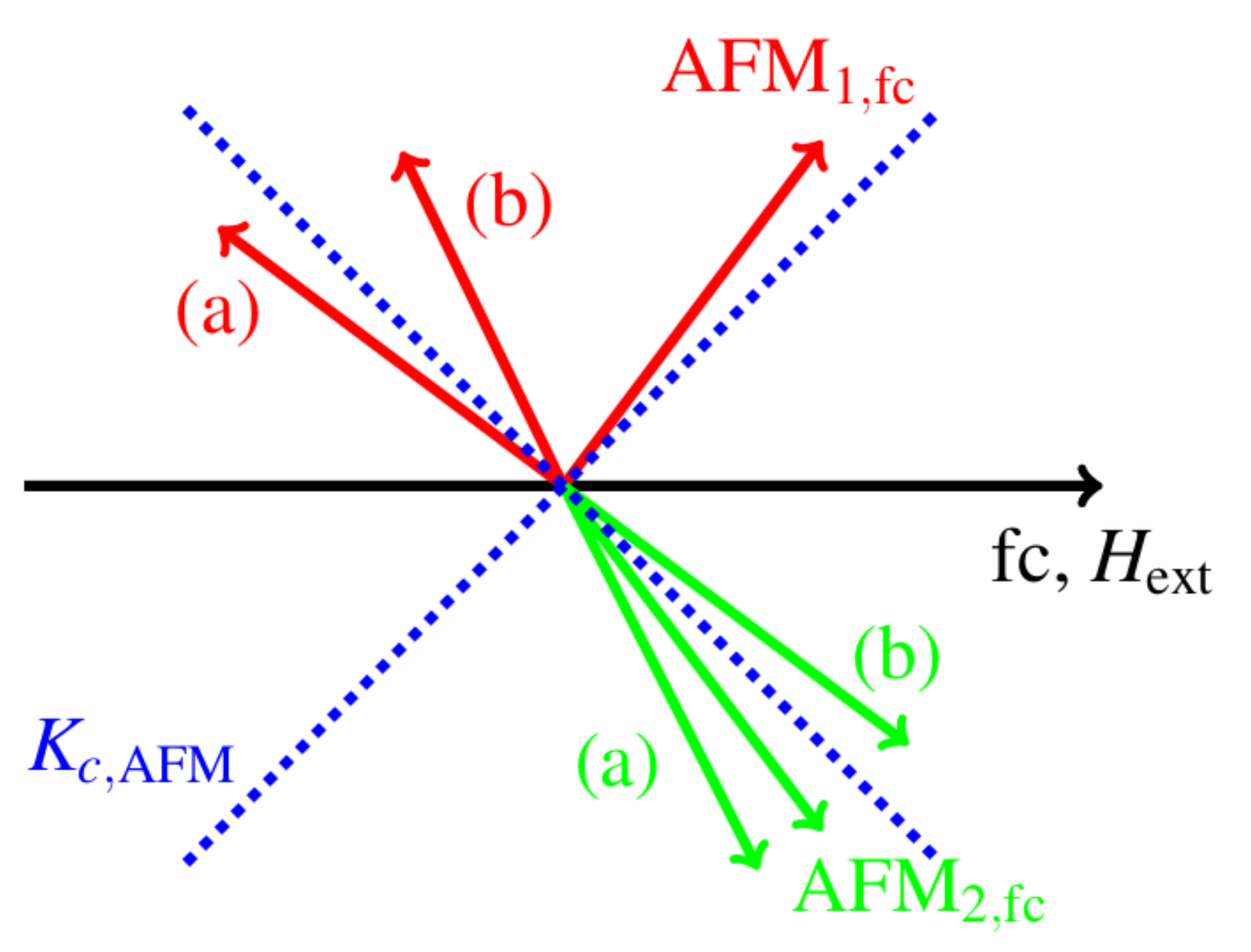}
\caption{Average behavior of an AFM grain, initialized in the spin flop state AFM$_\mathrm{1,fc}$ and AFM$_\mathrm{2,fc}$, whose easy axes $K_\mathrm{c,AFM}$ make an angle of 45$^{\circ}$ with respect to the field cooling direction $\mathrm{fc}$. The 2 sublattices (red and green) relax from a non collinear to the antiparallel state $(a)$ after the FM has switched towards negative saturation in $n = 1$. Position (b) represents the configuration of the AFM layers at positive saturation again.}
\label{fig:grain_Hoffmann} 
\end{figure}

\begin{figure}[htb]
  \centering
  \subfloat[field cooled]{\includegraphics[width=0.15\textwidth]{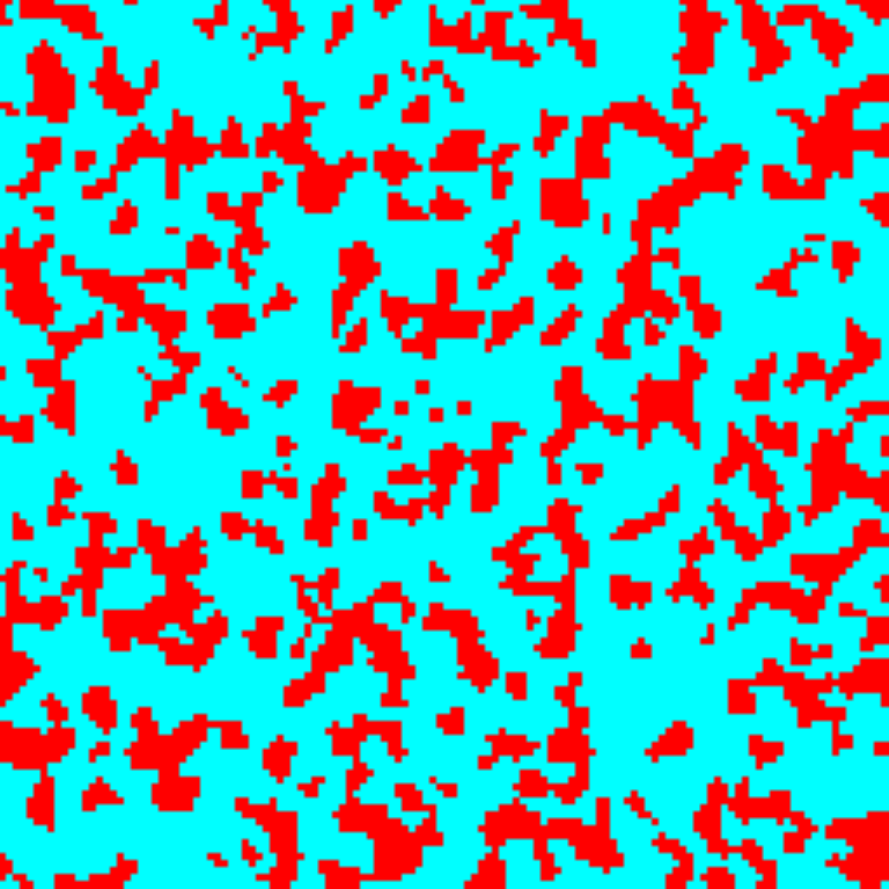}}
  \hfill
  \subfloat[negative saturation]{\includegraphics[width=0.15\textwidth]{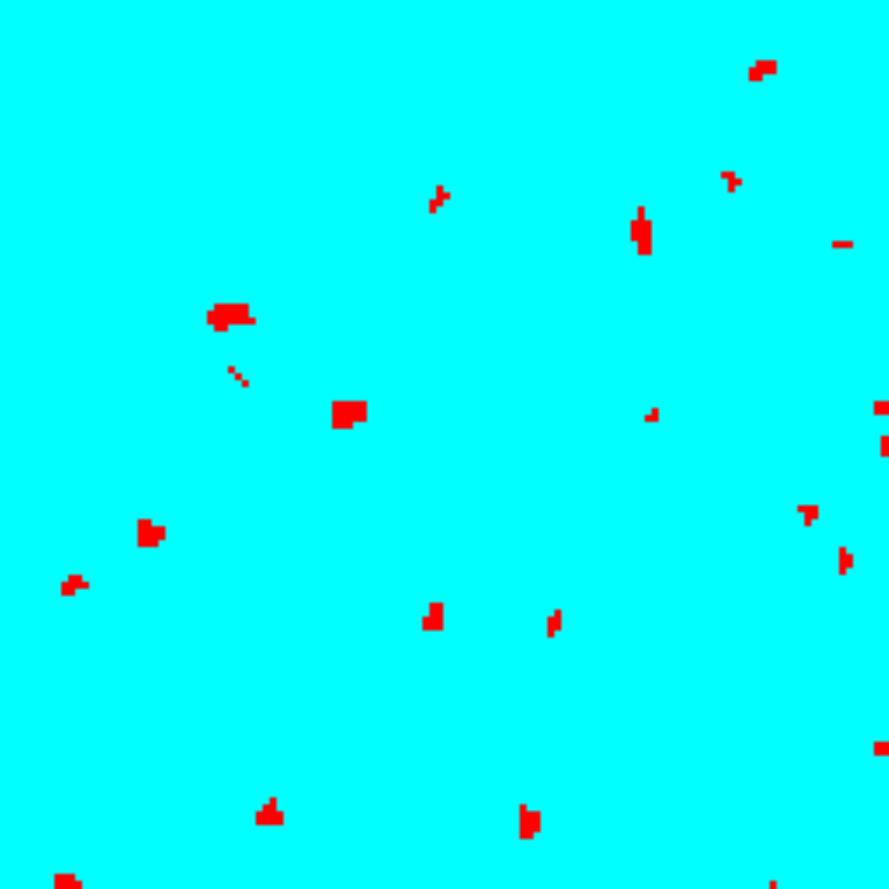}}
  \hfill
  \subfloat[positive saturation]{\includegraphics[width=0.15\textwidth]{AFM_IrMn_sat}}
  \caption{Training effect in an IrMn/CoFe bilayer. Red color: grains in which the 2 AFM layers are non collinear, blue color: grains in which the 2 AFM layers are antiparallel. Only the AFM layers in the pinned uncompensated grains stay parallel after reaching negative saturation in the first hysteresis loop. A quarter of the simulation box is shown.}
\label{fig:IrMn_AFM}
\end{figure}

\begin{figure}[htb]
\begin{center}
 \includegraphics[width=0.34\textwidth]{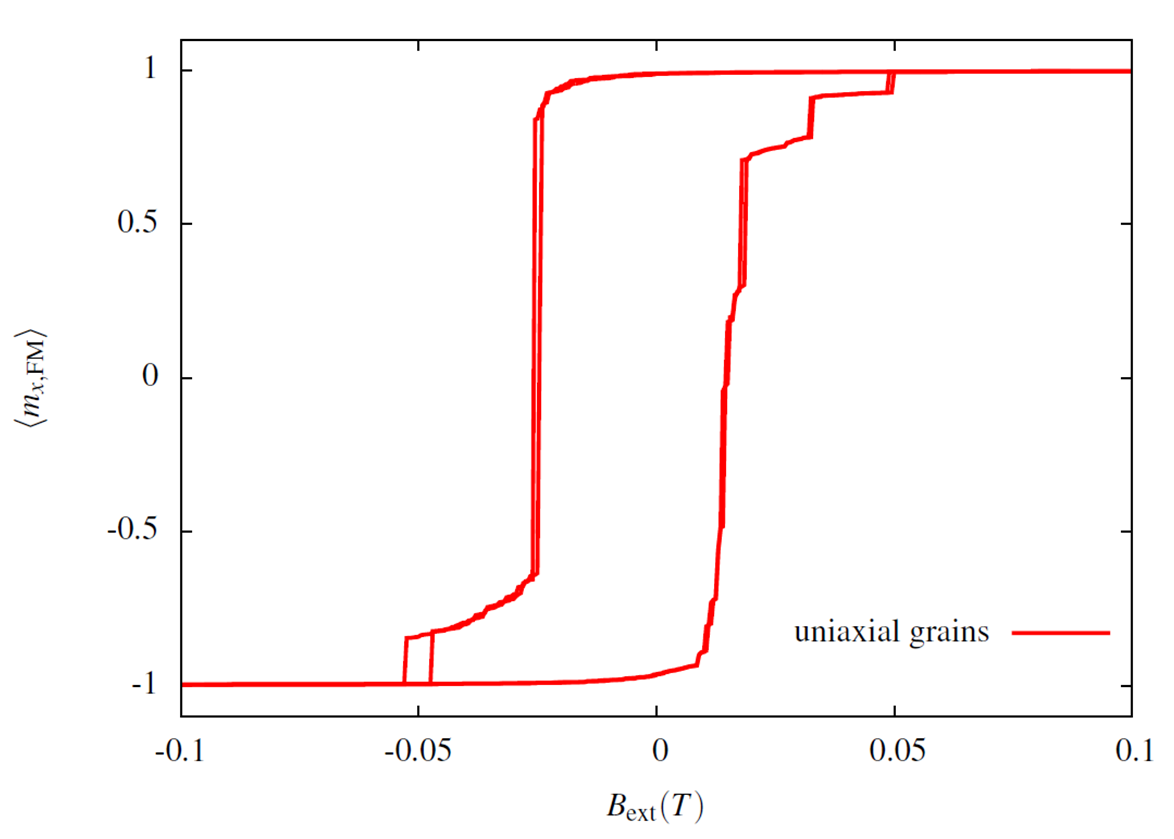}
\caption{Hysteresis loop for a polycrystalline AFM with uniaxial anisotropy $K_\mathrm{AFM} = \frac{K_\mathrm{c,AFM}}{4}$, due to the definition of the biaxial anisotropy energy density.}
\label{fig:uniaxial} 
\end{center}
\end{figure}

The simulation box is divided into 512 x 512 x ( 2 AFM + 1 FM ) cells of 3 nm lateral size and 10 nm in thickness. For the FM, typical parameters\cite{param_CoFe,param_CoFe2} were used: $M_\mathrm{FM}$ = 1600 kA/m and $A_\mathrm{FM} = 2.5 \times 10^{-11}$ J/m. A small uniaxial anisotropy ($K_\mathrm{FM}$ = 4 kJ/m$^3$) was applied in the field cooling direction to model magnetic annealing. The antiferromagnetic IrMn was divided into 20 nm grains using a Voronoi tessellation with randomly distributed anisotropy axes. In the AFM no intergrain interaction was taken into account. To simulate an infinite thin film, periodic boundary conditions were applied, using a macro geometry approach where 5 copies of the system were taken into account in the in-plane directions.\cite{MuMax3,pbc}
\\
The parameters\footnote{For a rescaling of the AFM parameters to achieve the correct total energy, see Supplementary Material.} of the AFM layer were tuned in order to match the experimental hysteresis loop: $A_\mathrm{AFM} = -1.88 \times 10^{-11}$ J/m, $A_\mathrm{I} = 1.7\times 10^{-11}$ J/m and as biaxial anisotropy constant we used $K_\mathrm{c,AFM} = 2.25 \times 10^5$ J/m$^3$. Approximately 20 \% of the AFM spins were initialized into a 45$^{\circ}$ canted configuration with respect to the field cooling direction and 76 \% were randomly distributed, with antiparallel sublattices. In order to produce a small exchange bias field for $n \geq 2$, about \mbox{4 \%} pinned uncompensated spins\footnote{As we have noted before one cannot distinguish between an uncompensated AFM spin or 2 compensated biaxial AFM spins, which are frozen into a 90$^\circ$ canted state, as both give rise to the same net effect.} were randomly added to the AFM layers, as discussed before (section \ref{sec:impl}). 
A quasistatic hysteresis loop was simulated along the field cooling direction. The result is shown in figure \ref{fig:CoFe}. 
One can clearly see the asymmetry and the reduction of the coercivity in the first hysteresis loop. During the hysteresis loop of the polycrystalline IrMn/CoFe bilayer, also the average behaviour of an AFM grain, whose easy axes $K_\mathrm{c,AFM}$ make an angle of 45$^{\circ}$ with respect to the field cooling ($\mathrm{fc}$) direction, was followed and is displayed in figure \ref{fig:grain_Hoffmann}. Starting from the spin flop initialized states AFM$_\mathrm{1,fc}$ and AFM$_\mathrm{2,fc}$, the AFM sublattices relax towards position $(a)$ when the FM reaches negative saturation. This relaxation from a non collinear to an antiparallel state produces athermal training as this is an irreversible transition. When the FM is saturated again in the field cooling direction, AFM$_1$ does not return to its initial position, but relaxes towards position $(b)$ and thus stays in an almost antiparallel state with AFM$_2$. In further hysteresis loops AFM$_1$ and AFM$_2$ only switch between positions $(a)$ and $(b)$. As this is a reversible transition, no training effect is obtained anymore for $n > 1$.
\\
In figure \ref{fig:IrMn_AFM} one can see in which grains the 2 AFM layers are in a non collinear (red color) and  antiparallel\footnote{The 2 AFM layers inside a grain are considered antiparallel when the angle between the magnetization in each is larger than 140$^{\circ}$ as in our case the angle between the 2 magnetization vectors will always be larger than 90$^{\circ}$, see e.g. the field cooled state in figure \ref{fig:grain_Hoffmann}.} (blue color) state after field cooling and at negative saturation for n = 1. One can see that after the first drop in the hysteresis loop towards negative saturation, the 2 AFM layers are antiparallel in all grains, expect in those of the pinned uncompensated AFM grains. For comparison, the case of a polycrystalline IrMn layer with uniaxial anisotropy is shown in figure \ref{fig:uniaxial}. As expected, no training effect is found.
\\

\section{Conclusions}
We have shown how to implement compensated antiferromagnetic interfaces to model hysteresis loops of ferromagnets using MuMax$^3$. This can be achieved by either using the small canting approximation in the case of low coupling constants or by adding two extra AFM layers to the simulation box. By using the latter micromagnetic approach, we can not only reproduce athermal training effects or spin flop coupling, but we can also take into account compensated and uncompensated AFM spins in one micromagnetic simulation and therefore give a good description of realistic FM/AFM interfaces. By demonstrating this model, we have opened a door for new micromagnetic studies of complex AFM/FM interfaces with both compensated and uncompensated spins. Static problems, such as the stability of the FM domain configurations can be investigated. This implementation can easily be extended to antiferromagnets with higher symmetries, to cases where the 2 sublattices couple differently towards the ferromagnet or to cases where the antiferromagnet consists of 3 sublattices.

\section{Acknowledgements}
This work was supported by the Flanders Research Foundation (FWO). J. Leliaert is supported by the Ghent University Special Research Fund (BOF). We gratefully acknowledge the support of NVIDIA Corporation with the donation of the Titan Xp GPU used for this research.

\section{References}

\bibliographystyle{model1-num-names}
\bibliography{article.bbl}

\clearpage


\section*{Supplementary material}

# Modelling compensated antiferromagnetic interfaces with MuMax$^3$

# Supplementary Material

Jonas De Clercq*, Jonathan Leliaert and Bartel Van Waeyenberge

DyNaMat, Department of Solid State Sciences, Ghent University
Krijgslaan 281 S1, B-9000 Gent, Belgium

## 1 Implementation of a compensated AFM/FM interface in MuMax$^3$

A compensated AFM with thickness $t_{\text{AFM}}$ can be modelled in MuMax$^3$[1] by adding 2 extra layers, labeled by AFM$_1$ and AFM$_2$, to the simulation box as shown in figure 1. In this way, the AFM layer thickness is equal to the cell size perpendicular to the FM/AFM interface, i.e. $t_{\text{AFM}} = C_z$. If $C_z \neq t_{\text{AFM}}$, the energy terms in the AFM layers have to be rescaled as detailed in section 1.6. To these two AFM layers one can attribute effective parameters, e.g. an anisotropy constant $K_{\text{AFM}}$, to ensure all the energy contributions of the AFM are included. As only nearest neighbouring cells are taken into account for the evaluation of the exchange energy in MuMax$^3$, one has to couple the bottom layer AFM$_1$ with the FM layer explicitly by adding a custom field and corresponding custom energy term.

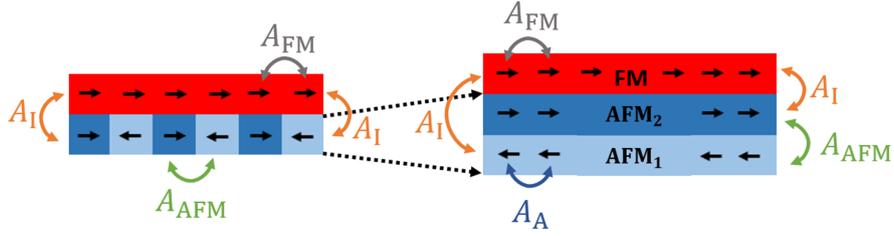

Figure 1: Representation of the micromagnetic model of a compensated AFM/FM in MuMax$^3$. AFM$_2$ can be coupled to the FM by rescaling the exchange energy. The FM and AFM$_1$ layer have to be coupled to each other using periodic boundary conditions or by defining a custom field / energy term.

### 1.1 AFM domain wall energy and physical interpretation of the interlayer and intralayer exchange stiffness

To demonstrate the validity of our micromagnetic model for the AFM, we investigate the energy of a 90° domain wall, shown in figure 2. It also illustrates the physical meaning of the interlayer exchange stiffness $A_{\text{AFM}} < 0$ and the intralayer exchange stiffness $A_A > 0$. In our micromagnetic model (figure 2, bottom), the atomic AFM sublattices are separated into 2 layers (AFM$_1$ and AFM$_2$) and the interlayer interaction $A_{\text{AFM}}$ replaces the negative exchange stiffness $A_1$ associated with the nearest neighbour interaction between two antiparallel AFM spins, e.g. spin 1 and 2 in figure 2 (top). The intralayer exchange stiffness $A_A$ of our model is responsible for the lateral coupling in the AFM spins, so it ensures a domain wall can be formed in the AFM.

To determine the correct value of the exchange stiffness $A_A$, we compare the domain wall energy of the two models (figure 2 (top) and figure 2 (bottom)), given by $\varepsilon_{\text{at}}$ and $\varepsilon_{\text{m}}$ respectively. Demanding that the total domain wall energy of $2N$ spins spread over a distance $2N\Delta$ is the same in both cases, one obtains

$$\begin{cases} \varepsilon_{\text{at}} = -(2N-1)\frac{A_{\text{AFM}}}{\Delta^2}\cos(\theta) \\ \varepsilon_{\text{m}} = -2(N-1)\frac{A_A}{(2\Delta)^2}\cos(\theta_m) \end{cases} \quad (1)$$

---
*Corresponding author: Jonas.DeClercq@ugent.be

1

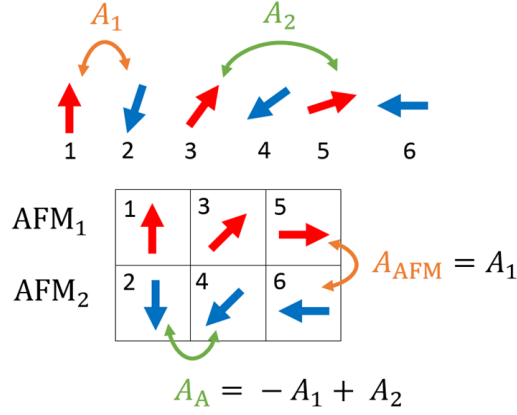

Figure 2: Top: representation of an AFM domain wall. $A_1$ and $A_2$ represent the exchange stiffnesses which can be associated with the nearest and next nearest neighbour interaction. Bottom: micromagnetic model of an AFM domain wall. AFM$_1$ and AFM$_2$ are the 2 AFM layers. $A_{\text{AFM}}$ and $A_A$ are the interlayer and intralayer exchange stiffnesses respectively. In our micromagnetic model spin 2 and 3 are not coupled.

with $\theta$ the angle between neighbouring AFM spins, at a distance $\Delta$ and $\theta_m$ the angle between two next nearest neighbouring AFM spins at a distance $2\Delta$. Introducing the angle $\alpha = \pi - \theta$, which will be small for large $N$, $\theta_m \approx 2\alpha$. Using the small angle approximation we find

$$\begin{cases} \varepsilon_{\text{at}} = (2N-1)\frac{A_{\text{AFM}}}{\Delta^2}\cos(\alpha) \approx (2N-1)\frac{A_{\text{AFM}}}{\Delta^2}\left(1-\frac{\alpha^2}{2}\right) \\ \varepsilon_m = -2(N-1)\frac{A_A}{4\Delta^2}\cos(2\alpha) \approx -(N-1)\frac{A_A}{2\Delta^2}\left(1-2\alpha^2\right) \end{cases} \quad (2)$$

So for large $N$ and neglecting constant energy offsets, we obtain that the total energy densities $\varepsilon_{\text{at}}$ and $\varepsilon_m$ are equal if $A_A \approx -A_{\text{AFM}}$.

If also the next nearest neighbour ferromagnetic exchange interaction $A_2$ is considered (e.g. between spins 3 and 5 in figure 2 (top)), it will need to be added to the intralayer exchange stiffness $A_A$. So in our micromagnetic model $A_A$ will generally be larger than $|A_{\text{AFM}}|$.

Another approach would be by correctly averaging out the interaction of a macrospin in 1 AFM layer with the next nearest neighbours of the other AFM layer by adding extra custom field and energy terms to our micromagnetic model. In this 1 dimensional model, spin 4 would interact with spin 3 and a weighted interaction with spin 5 and spin 1, to restore the translational symmetry of the exchange energy.

## 1.2 Setting the exchange coupling between the FM and AFM layers

When a FM layer with exchange stiffness $A_{\text{FM}}$ and saturation magnetisation $M_{\text{FM}}$ and an AFM layer with an intralayer exchange stiffness $A_A$ and saturation magnetisation $M_{\text{AFM}}$ are neighbours, MuMax$^3$ will average out the interlayer contribution to the exchange energy by calculating the harmonic mean $\alpha_H$ of the quantities $\frac{A_{\text{FM}}}{M_{\text{FM}}}$ and $\frac{A_A}{M_{\text{AFM}}}$ which is defined as

$$\alpha_H = \frac{2\frac{A_{\text{FM}}}{M_{\text{FM}}}\frac{A_A}{M_{\text{AFM}}}}{\frac{A_{\text{FM}}}{M_{\text{FM}}} + \frac{A_A}{M_{\text{AFM}}}} \quad (3)$$

This averaging follows from the fact that in MuMax$^3$ the exchange stiffness is attributed to the volume of a cell, rather than being defined at the interface between 2 cells.

To obtain a total surface energy density $J_I$ between the AFM$_2$ and the FM layer, one has to rescale the exchange energy density between the FM and AFM$_2$ regions by making use of the scaling factor $S$, as introduced in equation 9 of the MuMax$^3$ paper[1]. One can calculate that the scaling factor $S$ has to be given by

$$S = \frac{J_I C_z}{\alpha_H (M_{\text{FM}} + M_{\text{AFM}})} \quad (4)$$

This scaling factor has to be introduced in the function `ext_ScaleExchange` as defined in the MuMax$^3$ API which can be found at http://mumax.github.io/api.html.

## 1.3 Saturation magnetisation of the AFM layers

To divide the interface energy density $J_I$ equally between a FM and AFM cell, one has to set the saturation magnetisations (as micromagnetic input parameters in MuMax$^3$) of the ferromagnet and antiferromagnet to equal values, i.e. $M_{FM} = M_{AFM}$. As no demagnetisation energy is taken into account for the AFM layers, changing $M_{AFM}$ has no influence on the evolution to the minimal energy in the micromagnetic simulations. As the effective field terms acting on the AFM cell scale with $\frac{1}{M_{AFM}}$, the direction of the total effective field is not changed and thus still points in the direction of minimal energy. This rescaling only results in a rescaling of the time in the Landau-Lifshitz-Gilbert equation.

## 1.4 Coupling of the extra AFM layer to the FM layer by a custom energy term

To ensure that also the AFM layer, which is not directly neighbouring the FM layer, is exchange coupled to the FM layer a custom effective field and energy term[1] can be added in MuMax$^3$. The contribution to the exchange energy density[1] in a FM cell due to the coupling with an AFM$_1$ cell can be expressed as

$$\varepsilon_{ex,FM} = -\frac{1}{2} M_{FM} \vec{m}_{FM} \cdot \vec{B}_1 \qquad (5)$$

with $\vec{m}_{FM}$ the normalised magnetisation vector of the FM cell and the effective field term $\vec{B}_1$ as

$$\vec{B}_1 = \frac{J_I}{C_z M_{FM}} (\vec{m}_{AFM} - \vec{m}_{FM}) \qquad (6)$$

with $\vec{m}_{AFM}$ the normalised magnetisation vector of the underlying AFM$_1$ cell. The contribution to the energy density in an AFM$_1$ cell due to the coupling with a FM cell is

$$\varepsilon_{ex,AFM} = -\frac{1}{2} M_{FM} \vec{m}_{AFM} \cdot \vec{B}_2 = \varepsilon_{ex,FM} \qquad (7)$$

To ensure the interface energy density is equally divided between both layers, one has to set $M_{FM} = M_{AFM}$ as discussed in subsection 1.3. For the effective field in the AFM$_2$ cell, we can write

$$\vec{B}_2 = \frac{J_I}{C_z M_{FM}} (\vec{m}_{FM} - \vec{m}_{AFM}) = -\vec{B}_1 \qquad (8)$$

Practically, the coupling of AFM$_1$ with the FM layer can easily be implemented in MuMax$^3$ (from version 3.9.3) by adding following lines of code to the input script, assuming the FM is the top layer (n=2) and AFM$_1$ is the bottom layer (n=0).

```
f_I := (J_I)/(Cz*M_FM)}
B_1 := Madd( M, shifted(M,0,0,2), -f_I,f_I)
B_2 := Masked(Madd( M, shifted(M,0,0,-2), -f_I,f_I),layer(0))
B_ex := Add(Masked(B_1,layer(2)),B_2)
AddFieldTerm(B_ex)
AddEdensTerm(Mul(Mul(Const(-0.5),M_FM),Dot(M,B_ex)))
```

## 1.5 Establishing the intersublattice exchange coupling between the two AFM layers

The interaction between the 2 layers AFM$_1$ and AFM$_2$ (see figure 1), determined by the surface energy density $\delta t_{AFM}$, can be implemented by rescaling the exchange exchange stiffness between the 2 AFM layers, analogous to what was discussed in section 1.2. To obtain a total interface energy density $\delta t_{AFM}$, one can calculate that the scaling factor S has to be given by

$$S = -\frac{\delta t_{AFM} C_z}{2A_A} = \frac{A_{AFM}}{A_A} \qquad (9)$$

Note that the intralayer exchange stiffnesses $A_A$ in the layers AFM$_1$ and AFM$_2$ are equal as each layer represents an AFM sublattice. In general, the exchange stiffness $A_A$ can be different from the quantity $A_{AFM} < 0$ (see section 1.1). In our simulations $A_A$ is not a critical parameter as long as each layer inside an AFM grain is homogeneous and so we set $A_A = -A_{AFM}$, taking into account only the nearest neighbour interaction.

---

[1]If the FM exists of only 1 layer, one can also use periodic boundary conditions, in the direction perpendicular to the interface, i.e. `setPBC(0,0,1)`, and use the same scaling factor S as defined in section 1.2.

## 1.6 Rescaling energy terms

As MuMax³ uses a finite difference discretization, the simulation box has to be divided into cells of equal sizes. If the total thickness of a simulation box is given by $S_Z$ and is divided into $N_Z$ cells in the $z$-direction, then each cell has a thickness $C_z = \frac{S_z}{N_z}$. Generally, the thickness $t_{\text{FM}}$ of the FM is different from the thickness $t_{\text{AFM}}$ of the 2 AFM layers and a rescaling of the micromagnetic constants should be performed in order to obtain the correct total energy in the AFM layers. Choosing[2] a cell size $C_z$, perpendicular to the interface, the physical parameters $A_A$ and $K_{\text{c,AFM}}$ should be rescaled as

$$A_A \to A_A \left(\frac{t_{\text{AFM}}}{C_z}\right)$$

$$K_{\text{c,AFM}} \to K_{\text{c,AFM}} \left(\frac{t_{\text{AFM}}}{C_z}\right)$$

where the values on the right hand side are the MuMax³ input parameters. For example, considering an AFM layer with thickness $t_{\text{AFM}}$ and physical sublattice anisotropy constant $K_{\text{c,AFM}}$, we find that the total anisotropy energy[3] in an AFM cell with cell sizes $C_x$, $C_y$ and $C_z$ is then given by

$$E_K = \frac{K_{\text{c,AFM}}}{4}\left(\frac{t_{\text{AFM}}}{C_z}\right)C_xC_yC_z = \frac{K_{\text{c,AFM}}}{4}C_xC_yt_{\text{AFM}} \tag{10}$$

which amounts to the correct physical anisotropy energy in an AFM cell with thickness $t_{\text{AFM}}$ and lateral sizes $C_x$ and $C_y$.

Also the Zeeman energy of the AFM layers can be taken into account by appropriately rescaling its value

$$B_{\text{AFM}} \to B_{\text{AFM}} \left(\frac{t_{\text{AFM}}}{C_z}\right)$$

Note that in MuMax³, the saturation magnetisation $M_{\text{AFM}}$ of the antiferromagnet should be set equal to that of the ferromagnet, as discussed in section 1.3. One can take into account the correct Zeeman energy of the AFM layers by rescaling the magnitude of the external field with a factor $\frac{M_{\text{AFM,ph}}}{M_{\text{FM}}}$ where $M_{\text{AFM,ph}}$ represents the physical sublattice saturation magnetisation of an antiferromagnet.

## 2 Example

As an example of the implementation of a compensated AFM interface in MuMax³, we will simulate the hysteresis loop of a thin, infinite Ni$_{80}$Fe$_{20}$ layer ($t_{\text{FM}} = 3$ nm), coupled to an AFM with thickness of $t_{\text{AFM}} = 3$ nm, uniaxial sublattice anisotropy constant $K_{\text{AFM}} = 7 \times 10^5$ J/m³, interface coupling $J_I = 0.7$ mJ/m² and mutual interaction $\delta = 1 \times 10^6$ J/m³ between the 2 AFM layers. To compare our results with the macrospin approach, the demagnetizing field is also switched off in the FM. Because we cannot model the dynamics of the AFM layer (section 1.1), only the `relax()` function can be used to minimize the energy of the system. This solver switches off precession in contrast to the `run()` function.

As the FM is discretized in only 1 layer and thus $t_{\text{FM}} = t_{\text{AFM}} = C_z$, it is sufficient to couple AFM$_1$ with the FM by making use of periodic boundary conditions in the $z$-direction and thus exchange couple the bottom layer (n=0) with the top layer (n=2). When the FM is discretized into multiple layers, one has to leave out the statements `ext_ScaleExchange(0, 2, J_I*Cz/(2*M_FM*alpha_H))` and `setPBC(0,0,1)` in the code and add the custom code to the input script, as discussed in section 1.4. The FM was initialized in the spin flop state and the external field applied perpendicular to the anisotropy axis of the AFM. A small angle was introduced to break the symmetry of the system.

```
Nx := 64
Ny := 64
Nz := 3
Cx := 3.0e-9
Cy := 3.0e-9
Cz := 3.0e-9
```

---

[2]If we discretize the FM in only 1 layer, we will set $C_z = t_{\text{FM}}$.

[3]In case the magnetisation vector of the AFM cell is aligned at an angle of 45° with respect to the biaxial anisotropy axes.

```
setgridsize(Nx, Ny, Nz)
setcellsize(Cx, Cy, Cz)
setgeom(universe())
setPBC(0,0,1) // as the FM is discretized in only 1 layer

DefRegion(0, Layer(0)) // AFM1 layer
DefRegion(1, Layer(1)) // AFM2 layer
DefRegion(2, Layer(2)) // FM layer

A_FM := 1.3e-11
M_FM:= 800e3
Aex.setRegion(2, A_FM)
Msat.setRegion(2, M_FM)
NoDemagSpins.SetRegion(2, 1) // only to compare with the macrospin approach
m.setRegion(2, uniform(1, 0, 0))

t_AFM := Cz
M_AFM := M_FM // to equally divide the energy density between a FM and AFM cell
K_AFM := 7.0e5
J_I := 0.7e-3
delta := 1.0e6
A_AFM := - delta*t_AFM*Cz/2
A_A := -A_AFM
for i:=0; i<2; i++ {
   NoDemagSpins.SetRegion(i, 1) // no demagnetisation energy for AFM layers
   Aex.setRegion(i, A_A)
   Msat.setRegion(i, M_AFM)
   m.setRegion(i, uniform(0.1, pow(-1,i), 0)) // sublattices are initialized antiparallel
   Ku1.setregion(i,K_AFM)
   anisU.setregion(i,vector(0,1,0))
}

alpha_H := (2/M_FM)*(1/(1/A_FM + 1/A_A)) // harmonic mean for M_FM = M_AFM
ext_ScaleExchange(1, 2, J_I*Cz/(2*M_FM*alpha_H)) // rescaling coupling between AFM2 and FM
ext_ScaleExchange(0, 2, J_I*Cz/(2*M_FM*alpha_H)) // rescaling coupling between AFM1 and FM
ext_ScaleExchange(0, 1, A_AFM/A_A) // rescaling coupling between AFM1 and AFM2

relax()
save(m)

B_app := 0.0
TableAddVar(B_app, "B_app", "T") // save external field applied on FM layer
TableAdd(m.Region(2)) // save average magnetization of FM layer
for i := 80; i > -81; i-- {
   B_app = i * 1e-3
   B_ext.setregion(2, vector(B_app*cos(1.0*pi/180), B_app*sin(1.0*pi/180), 0))
   tablesave()
   relax()
}
for i := -79; i < 81; i++ {
   B_app = i * 1e-3
   B_ext.setregion(2, vector(B_app*cos(1.0*pi/180), B_app*sin(1.0*pi/180), 0))
   tablesave()
   relax()
}
```

Using the coercive fields of the hysteresis loop (see figure 3), we find that $B_c \approx 40$ mT which is in correspondence with equation 9 of the paper for $K_{FM} = 0$.

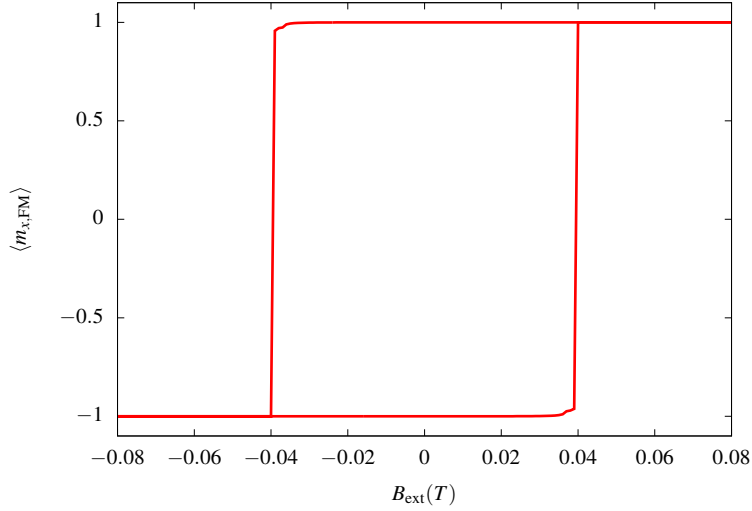

Figure 3: Hysteresis loop of the FM layer, produced by using this example.

## 3 Implementation of biaxial anisotropy

Although biaxial anisotropy is not implemented natively in MuMax$^3$, it can be included by adding a custom field and corresponding custom energy term. The biaxial anisotropy energy density is given by

$$\varepsilon_K = K_\text{c}(\vec{c}_1.\vec{m})^2(\vec{c}_2.\vec{m})^2 \tag{11}$$
$$= K_\text{c}\cos^2(\beta)\sin^2(\beta) \tag{12}$$

where $K_\text{c}$ is the biaxial anisotropy constant, $\vec{c}_1$ and $\vec{c}_2$ are the *normalised* biaxial anisotropy axes and $\beta$ is the angle that the normalised magnetisation vector $\vec{m}$ makes with respect to $\vec{c}_1$.

If we take e.g. $\vec{c}_1$ and $\vec{c}_2$ along the *x* and *y* direction of the simulation box respectively, then one can add following code to the MuMax$^3$ input script

```
c1 := Constvector(1,0,0)
c2 := Constvector(0,1,0)
f := Const (-2 * Kc / (M_FM))
B_c := Mul(f, Madd(Mul( Mul(Mul( Dot(c2, m)
        , Dot(c2, m)),Dot(c1, m)),c1)
        , Mul( Mul(Mul( Dot(c1, m)
        , Dot(c1, m)),Dot(c2, m)),c2), 1,1))
AddFieldTerm(B_c)
AddEdensTerm(Mul(Mul(Const(-0.25),Const(M_FM)),Dot(M,B_c)))
```

to implement this biaxial anisotropy. Note that in this case the maximal energy density is obtained for $\varepsilon_K = \frac{K_\text{c}}{4}$ at the angle $\beta = \frac{\pi}{4}$.

## 4 Spin flop coupling for biaxial antiferromagnets

In this section we will show that the expression for the spin flop coupling in an antiferromagnet with biaxial anisotropy, is the same for uniaxial antiferromagnets (after Hoffmann training). Assuming that the 2 antiferromagnetic macrospins are in an almost antiparallel position, one can define the surface energy density $\sigma_\text{AFM}$ as

$$\sigma_\text{AFM} = -J_\text{I}\cos\left(\frac{3\pi}{4} - \theta - \beta\right) - J_\text{I}\cos\left(\frac{\pi}{4} - \phi + \beta\right) - \delta t_\text{AFM}\cos\left(\theta + \phi\right)$$
$$- \frac{K_\text{c,AFM}t_\text{AFM}}{4}\sin^2\left(2\theta + \frac{3\pi}{2}\right) - \frac{K_\text{c,AFM}t_\text{AFM}}{4}\sin^2\left(2\phi + \frac{\pi}{2}\right)$$

6

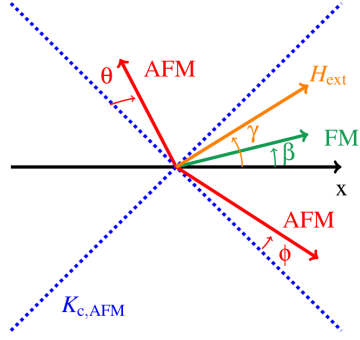

Figure 4: Definition of the AFM canting angles θ and φ, the FM angle β and the angle γ of the external field $H_{\text{ext}}$.

The definition of the angles θ, φ, γ and β are shown in figure 4. Using the small canting angle approximation, one finds that

$$\sigma_{\text{AFM}} = -J_I\left[1 - \frac{\theta^2}{2}\right]\cos\left(\frac{3\pi}{4} - \beta\right) - J_I\theta\sin\left(\frac{3\pi}{4} - \beta\right)$$

$$- J_I\left[1 - \frac{\phi^2}{2}\right]\cos\left(\frac{\pi}{4} + \beta\right) - J_I\phi\sin\left(\frac{\pi}{4} + \beta\right)$$

$$+ \frac{\delta t_{\text{AFM}}}{2}(\theta + \phi)^2 + K_{\text{c,AFM}}t_{\text{AFM}}\left(\theta^2 + \phi^2\right)$$

while dropping constant energy terms. By minimizing the energy density $\sigma_A$ towards the AFM angles θ and φ, and eliminating θ from the φ equation and vice versa, one obtains

$$\theta(\beta) = \frac{J_I\sin\left(\frac{\pi}{4} + \beta\right)\left[2K_{\text{c,AFM}}t_{\text{AFM}} + J_I\cos\left(\frac{\pi}{4} + \beta\right)\right]}{4K_{\text{c,AFM}}^2 t_{\text{AFM}}^2 + 4t_{\text{AFM}}^2 K_{\text{c,A}}\delta - J_I^2\cos^2\left(\frac{\pi}{4} + \beta\right)}$$

$$\phi(\beta) = \frac{J_I\sin\left(\frac{\pi}{4} + \beta\right)\left[2K_{\text{c,AFM}}t_{\text{AFM}} - J_I\cos\left(\frac{\pi}{4} + \beta\right)\right]}{4K_{\text{c,AFM}}^2 t_{\text{AFM}}^2 + 4t_{\text{AFM}}^2 K_{\text{c,AFM}}\delta - J_I^2\cos^2\left(\frac{\pi}{4} + \beta\right)}$$

Resubstituting these expressions in the energy density and retaining the lowest order approximation, one finally obtains that

$$\sigma_{\text{AFM}}(\beta) = \left[\frac{J_I^2}{2t_{\text{AFM}}(K_{\text{c,AFM}} + \delta)}\right]\cos^2\left(\frac{\pi}{4} + \beta\right)$$

This equation formally looks the same as the one for an antiferromagnet with uniaxial anisotropy, but with the easy axis rotated over 45°. The energy is minimal for $\beta = \frac{\pi}{4}$ or $\frac{5\pi}{4}$ and the hysteresis loops of the FM layer are analogous to the ones as defined in the Stoner Wohlfarth model with an anisotropy axis rotated over 45° with respect to the direction of the external field and so the switching field $B_c$ is reduced by a factor 2 for γ = 0, i.e.

$$B_c = \left[\frac{J_I^2}{2M_{\text{FM}}t_{\text{FM}}t_{\text{AFM}}(K_{\text{c,AFM}} + \delta)}\right] \quad (13)$$

neglecting the anisotropy of the ferromagnet.

## 5 Exchange bias for biaxial antiferromagnets

Analogous to the calculation of the coercivity induced by spin flop coupling in an antiferromagnet with uniaxial anisotropy, one can calculate the exchange bias field for an antiferromagnet with a strong biaxial anisotropy and low coupling constant. In this case the function $\sigma_{\text{AFM}}(\beta, \theta, \phi)$ is given by

$$\sigma_{\text{AFM}} = -J_I\cos\left(\frac{\pi}{4} + \theta - \beta\right) - J_I\cos\left(\frac{\pi}{4} + \phi + \beta\right)$$

$$- \delta t_{\text{AFM}}\cos\left(\theta + \phi - \frac{\pi}{2}\right) - \frac{K_{\text{c,AFM}}t_{\text{AFM}}}{4}\sin^2\left(2\theta + \frac{\pi}{2}\right)$$

$$- \frac{K_{\text{c,AFM}}t_{\text{AFM}}}{4}\sin^2\left(2\phi + \frac{\pi}{2}\right)$$

7

where we have defined the in-plane biaxial anisotropy as in section 3, but rotated over 45° with respect to the *x*-axis. The definition of the angles θ, φ, γ and β is shown in figure 5.

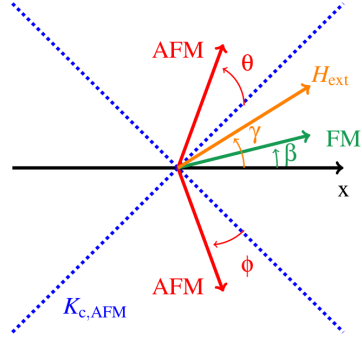

Figure 5: Definition of the AFM canting angles θ and φ, the FM angle β and the angle γ of the external field $H_{\text{ext}}$.

Using the small canting angle approximation again, one finds that

$$\sigma_{\text{AFM}} = -J_I \left[ \cos\left(\frac{\pi}{4} - \beta\right)\left(1 - \frac{\theta^2}{2}\right) - \theta \sin\left(\frac{\pi}{4} - \beta\right) \right]$$
$$- J_I \left[ \cos\left(\frac{\pi}{4} + \beta\right)\left(1 - \frac{\phi^2}{2}\right) - \phi \sin\left(\frac{\pi}{4} + \beta\right) \right]$$
$$- \delta t_{\text{AFM}}(\theta + \phi) - \frac{K_{\text{c,AFM}} t_{\text{AFM}}}{4}(1 - 4\theta^2)$$
$$- \frac{K_{\text{c,AFM}} t_{\text{AFM}}}{4}(1 - 4\phi^2)$$

Calculating the first derivative of $\sigma_{\text{AFM}}$ towards θ and φ to minimize $\sigma_{\text{AFM}}$, one obtains

$$\phi = -\frac{J_I \sin\left(\frac{\pi}{4} + \beta\right) - \delta t_{\text{AFM}}}{2K_{\text{c,AFM}} t_{\text{AFM}} + J_I \cos\left(\frac{\pi}{4} + \beta\right)} \tag{14}$$

$$\theta = -\frac{J_I \sin\left(\frac{\pi}{4} - \beta\right) - \delta t_{\text{AFM}}}{2K_{\text{c,AFM}} t_{\text{AFM}} + J_I \cos\left(\frac{\pi}{4} - \beta\right)} \tag{15}$$

and by resubstituting these 2 equation into $\sigma_{\text{AFM}}$

$$\sigma_{\text{AFM}}(\beta) = -\sqrt{2} J_I \cos(\beta) - \frac{1}{2}\left[\frac{\left(J_I \sin\left(\frac{\pi}{4} - \beta\right) - \delta t_{\text{AFM}}\right)^2}{2K_{\text{c,AFM}} t_{\text{AFM}} + J_I \cos\left(\frac{\pi}{4} - \beta\right)}\right] \tag{16}$$

$$- \frac{1}{2}\left[\frac{\left(J_I \sin\left(\frac{\pi}{4} + \beta\right) - \delta t_{\text{AFM}}\right)^2}{2K_{\text{c,AFM}} t_A + J_I \cos\left(\frac{\pi}{4} + \beta\right)}\right] \tag{17}$$

dropping constant terms. Neglecting the anisotropy in the FM, one can thus write the total surface energy density as

$$\sigma(\beta) = -\mu_0 M_{\text{FM}} t_{\text{FM}} H_{\text{ext}} \cos(\gamma - \beta) - \sqrt{2} J_I \cos(\beta) \tag{18}$$

$$- \frac{1}{2}\left[\frac{\left(J_I \sin\left(\frac{\pi}{4} - \beta\right) - \delta t_{\text{AFM}}\right)^2}{2K_{\text{c,AFM}} t_{\text{AFM}} + J_I \cos\left(\frac{\pi}{4} - \beta\right)}\right] \tag{19}$$

$$- \frac{1}{2}\left[\frac{\left(J_I \sin\left(\frac{\pi}{4} + \beta\right) - \delta t_{\text{AFM}}\right)^2}{2K_{\text{c,AFM}} t_{\text{AFM}} + J_I \cos\left(\frac{\pi}{4} + \beta\right)}\right] \tag{20}$$

For γ = 0, one finds after minimizing the effective surface energy density σ that β = 0 and β = π are 2 extrema, representing the saturated states of the FM. Numerical analysis shows that there are no other extrema

in our parameter range.[4] This implies that one cannot not define the bias field as $B_{eb} = B_{ext}(\langle m_{FM,x}\rangle = 0)$. One finds that the solution $\beta = 0$ is stable for the magnetic field

$$B_1 \geq -\frac{\sqrt{2}J_I}{M_{FM}t_{FM}}$$
$$+ \frac{1}{M_{FM}t_{FM}}\left[\frac{J_I^2}{2K_{c,AFM}t_{AFM} + \frac{\sqrt{2}}{2}J_I} + 2\frac{\left(\frac{\sqrt{2}}{2}J_I - \delta t_{AFM}\right)J_I^2}{\left(2K_{c,AFM}t_{AFM} + \frac{\sqrt{2}}{2}J_I\right)^2}\right]$$
$$- \frac{1}{M_{FM}t_{FM}}\left[\frac{\left(\frac{\sqrt{2}}{2}J_I - \delta t_{AFM}\right)J_I\sqrt{2}}{2K_{c,AFM}t_{AFM} + \frac{\sqrt{2}}{2}J_I} - \frac{\left(\frac{\sqrt{2}}{2}J_I - \delta t_{AFM}\right)^2 J_I^2}{\left(2K_{c,AFM}t_{AFM} + \frac{\sqrt{2}}{2}J_I\right)^3}\right]$$
$$+ \frac{1}{2M_{FM}t_{FM}}\frac{\left(\frac{\sqrt{2}}{2}J_I - \delta t_{AFM}\right)^2 J_I\sqrt{2}}{\left(2K_{c,AFM}t_{AFM} + \frac{\sqrt{2}}{2}J_I\right)^2}$$

and $\beta = \pi$ is stable for

$$B_2 \leq -\frac{\sqrt{2}J_I}{M_{FM}t_{FM}}$$
$$+ \frac{1}{M_{FM}t_{FM}}\left[-\frac{J_I^2}{2K_{c,AFM}t_{AFM} - \frac{\sqrt{2}}{2}J_I} + 2\frac{\left(\frac{\sqrt{2}}{2}J_I + \delta t_{AFM}\right)J_I^2}{\left(2K_{c,AFM}t_{AFM} - \frac{\sqrt{2}}{2}J_I\right)^2}\right]$$
$$+ \frac{1}{M_{FM}t_{FM}}\left[\frac{\left(\frac{\sqrt{2}}{2}J_I + \delta t_{AFM}\right)J_I\sqrt{2}}{2K_{c,AFM}t_{AFM} - \frac{\sqrt{2}}{2}J_I} - \frac{\left(\frac{\sqrt{2}}{2}J_I + \delta t_{AFM}\right)^2 J_I^2}{\left(2K_{c,AFM}t_{AFM} - \frac{\sqrt{2}}{2}J_I\right)^3}\right]$$
$$+ \frac{1}{2M_{FM}t_{FM}}\frac{\left(\frac{\sqrt{2}}{2}J_I + \delta t_{AFM}\right)^2 J_I\sqrt{2}}{\left(2K_{c,AFM}t_{AFM} - \frac{\sqrt{2}}{2}J_I\right)^2}$$

Using the symmetry of the hysteresis loop, one can define the exchange bias field $B_{eb}$ as the shift of the hysteresis loop, determined by $B_{eb} = \frac{B_1 + B_2}{2}$. Given fixed parameters $K_{c,AFM}$ and $\delta$, one can plot the bias field $B_{eb}$ as a function of the coupling parameter $J_I$. (see the main paper for the reduced anisotropy constant $k = 5$) For a very high anisotropy $K_{c,AFM}$, the bias field is reduced to that of 2 uncompensated spins, making an angle of 45° with the field cooling direction.

Numerical analysis of the effective surface energy density $\sigma(\beta)$ shows that there is a region in which both the solutions $\beta = 0$ and $\beta = \pi$ are stable (see figure 6), which results in a small coercivity $B_c$, given by

$$B_c = \frac{B_1 - B_2}{2}$$
$$= \frac{256 t_{AFM} J_I^6 \left(K_{c,AFM} + \frac{\delta}{4} - \frac{4K_{c,AFM}^3 t_{AFM}^2}{J_I^2}\right)\left(\frac{4K_{c,AFM}\delta t_{AFM}^2}{J_I^2} + 1\right)}{M_{FM}t_{FM}\left(\sqrt{2}J_I - 4K_{c,AFM}t_{AFM}\right)^3 \left(\sqrt{2}J_I + 4K_{c,AFM}t_{AFM}\right)^3}$$

as there are no other energy minima in our parameter range.

An example of the surface energy density for $j = 1$, $k = 5$ and external field $b_{ext} = -1.25$ is shown in figure 7. One can clearly see that for, these parameters, both the solutions $\beta = 0$ and $\beta = \pi$ are stable. For low coupling constants $J_I$, one finds that $B_c \ll B_{eb}$, as expected. At breakdown of the low canting angle approximation ($j \approx 1.5$) we obtain that $B_c \approx 5\%$ of $B_{eb}$ for $k = 5$. The coercivity $b_c$ as a function of the coupling constant $j$ is shown in figure 8.

---

[4]In the case of higher coupling constants $J_I$ there is a breakdown of our small canting angle approximation as can be seen in figure 13 of the main paper.

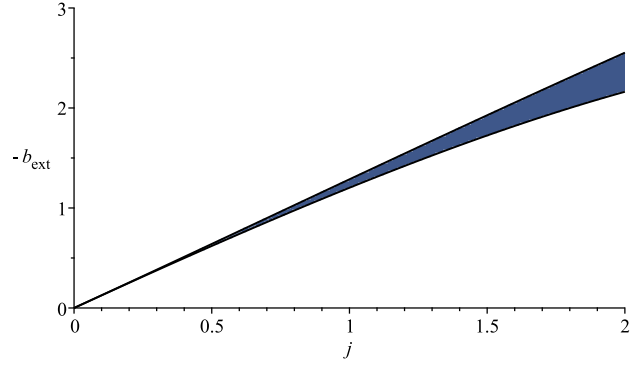

Figure 6: The 2 critical stability curves (black lines) determined by $\left(\frac{\partial^2 \sigma}{\partial \beta^2}\right) = 0$ for $\beta = 0$ and $\beta = \pi$ with $k = 5$ (reduced units). The blue area shows the stability overlap. Remark that a breakdown of the low canting angle approximation happens around $j = 1.5$.

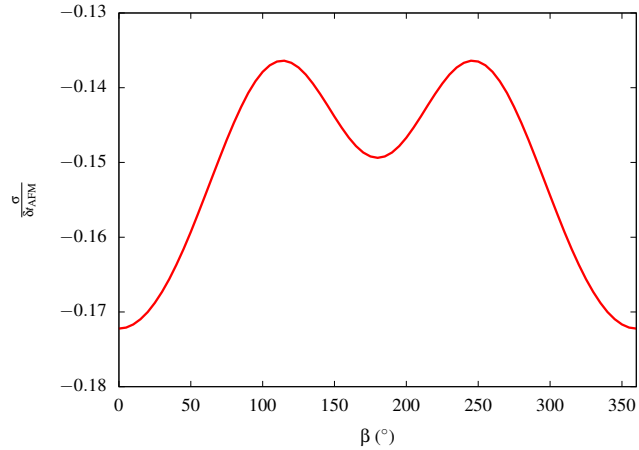

Figure 7: Reduced surface energy density $\frac{\sigma}{\delta t_{\text{AFM}}}$ as a function of the FM angle $\beta$ for the parameters $j = 1$, $k = 5$ and external field $b_{\text{ext}} = -1.25$. Two minima are present $\beta = 0$ (absolute minimum) and $\beta = \pi$ (relative minimum) in the region of stability overlap. This leads to a finite coercivity, in agreement with figure 6.

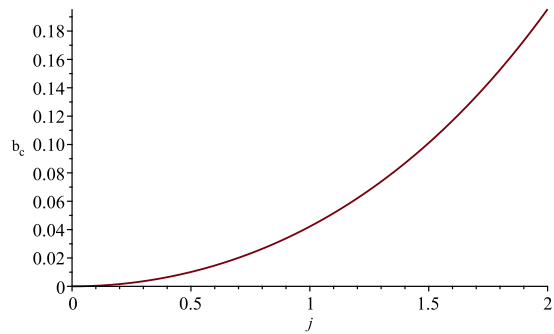

Figure 8: Coercivity $b_c$ as function of the coupling constant $j$ for $k = 5$ (reduced units).

The MuMax[3] API can be found at http://mumax.github.io/api.html



\end{document}